\documentclass[12pt]{article}

\DeclareSymbolFont{extraup}{U}{zavm}{m}{n}
\DeclareMathSymbol{\vardiamond}{\mathalpha}{extraup}{87}
\usepackage{color}
\definecolor{grey1}{gray}{0.4}
\definecolor{grey2}{gray}{0.7}
\definecolor{grey3}{gray}{0.81}
\definecolor{grey4}{gray}{0.87}
\definecolor{violet}{rgb}{0.95, 0, 1}
\definecolor{brown}{rgb}{0.59, 0.29, 0.0}
\definecolor{deepgreen}{rgb}{0.0, 0.5, 0.0}
\usepackage{scicite}
 \usepackage{hyperref}

\usepackage{times}

\usepackage{amsfonts,amssymb,graphicx}
\usepackage{amsthm}
\usepackage{amsmath}

\newenvironment{sciabstract}{%
\begin{quote} \bf}
{\end{quote}}

\newcounter{lastnote}

\title{ Egalitarianism in the rank aggregation problem:\\ a new dimension for democracy}

\author{
Pierluigi Contucci,$^{1}$ Emanuele Panizzi,$^{2}$ Federico Ricci-Tersenghi$^{3}$ and Alina S\^irbu$^{4}$\\ \\
\normalsize{$^{1}$Department of Mathematics, Alma Mater Studiorum, University of Bologna}\\
\normalsize{Piazza di Porta San Donato 5, 40126  Bologna, Italy}\\
\normalsize{$^{2}$Dipartimento di Informatica,  Sapienza Universit\`a di Roma,}\\ \normalsize{Via Salaria 113, 00198 Roma, Italy}\\
\normalsize{$^{3}$Dipartimento di Fisica, INFN -- Sezione di Roma 1 and CNR -- IPCF, UOS di Roma}\\
\normalsize{Sapienza Universit\`{a} di Roma, P.le A.~Moro 5, Roma 00185, Italy}\\
\normalsize{$^{4}$Department of Computer Science and Engineering, University of Bologna,}\\ 
\normalsize{Via Mura Anteo Zamboni 7, 40126 Bologna, Italy}\\
}

\date{}

\begin{document} 


\baselineskip 24pt


\maketitle 


\begin{sciabstract}
Winner selection by majority, in an election between two candidates,
is the only rule compatible with democratic principles. Instead, when the candidates 
are three or more and the voters rank candidates in order of preference, 
there are no univocal criteria for the selection of the winning (\emph{consensus}) ranking 
and the outcome is known to depend sensibly on the adopted rule.
Building upon XVIII century Condorcet theory, whose idea was to maximize 
total voter satisfaction, we propose here the addition of a new basic principle (\emph{dimension}) to guide the selection: satisfaction should be distributed among voters as equally as possible. 
With this new criterion we identify an optimal set of rankings. They range from the Condorcet 
solution to the one which is the most egalitarian with respect to the voters. We show that 
highly egalitarian rankings have the important property to be more stable with respect to fluctuations 
 and that classical consensus rankings (Copeland, Tideman, Schulze) 
often turn out to be non optimal. 
The new dimension we have introduced provides, when used together with 
that of Condorcet, a clear classification of all the possible rankings. 
By increasing awareness in selecting a consensus ranking our method
may lead to social choices which are more egalitarian compared 
to those achieved by presently available voting systems.
\end{sciabstract}


\paragraph*{Introduction}
A voting process starts with individuals giving a formal indication
of a choice (\emph{ballot}) or, more generally, a set of preferences between two or more candidates 
(or alternatives)\footnote{For the sake of simplicity, we prefer not to discuss partial rankings, 
because the meaning of not ranking a candidate may change a lot from application to application.}. The process ends 
with an aggregation procedure (\emph{winner selection method}) of these indications, in order to produce 
the \emph{consensus} ranking, that is the ranking on which voters should agree more upon and which should be the output of the election.
The complexity of the selection process comes, in general, from the presence of competing interests 
and conflicting opinions which make it impossible to satisfy all the preferences expressed by the voters.
With his seminal work on voting theory, Condorcet discovered \cite{condorcet} that the majority rule, applied to pairwise
preferences, may lead to invalid solutions. For instance in an election among three candidates the 
preferences may sum up to prefer the first to the second, the second to the third and the third to the first. 
Similarly, from the formal logic perspective, Arrow's theorem 
\cite{arrow} states that a perfectly fair voting system can not exist.
The lack of an ideal voting system in case of more than two candidates implies that any 
winner selection procedure contains some kind of arbitrariness and makes the studies
on voting methods an interesting research problem.

Typical examples of voting processes are political elections \cite{bookVoting}. In that case 
the need of a single winner, or a single winning ranking, has encouraged the use of very elementary
selection rules, easy to compute and to understand by voters and competitors at the
expense of making sub-optimal choices. 
Voting theory include also cases beyond political matters. Survey rankings for instance, typically made for commercial
purposes, like hotel listings, movie rankings, best product on the market etc, are selected 
with totally different criteria. The choice of the ten best smartphones, say, is not made
by maximizing the voters total satisfaction, but rather to ensure that each customer finds,
among those ten, a satisfactory model.

A similar problem is very much studied in computer science under the name of Rank Aggregation: 
a typical example is the merging of webpage rankings produced by different search engines or 
obtained according to different criteria \cite{rankAggr}. The main difference from the examples above is that here the 
number of voters (engines/criteria) is small, while the number of candidates (webpages) is large. 
This is why in that field of research the focus is more on the algorithmic challenge of computing 
the consensus ranking efficiently. Here we are more interested in presenting the new criterion for 
better selecting the consensus ranking; thus we concentrate on small number of candidates, so 
that all possible rankings (with ties) can be easily computed.

It is therefore clear that the problem of finding a good consensus ranking is an interdisciplinary 
topic of research: it is inspired and guided by studies in sociology, marketing, economy and 
political sciences. The disciplines technically involved in the solutions are statistics, 
mathematics and computer science.

\paragraph*{Definition of the problem}

Each of $n$ voters expresses a preference about $m$ candidates by sorting them 
in a ranked list, possibly with ties, resulting in $n$ ballots. Valid ranked lists for $m=4$ candidates are for instance
$\mathsf{B\!>\!A\!>\!D\!>\!C}$, $\mathsf{D\!>\!A\!=\!C\!>\!B}$ and $\mathsf{C\!>\!A\!=\!B\!=\!D}$. 
We call $\mathbf{r}_v$ the ballot of voter $v$. 
Each voter wishes the consensus ranking 
to be as close as possible to his ballot and, following Condorcet \cite{condorcet},
a good winner selection method should work by maximizing the total sum of those wishes, i.e. minimizing the 
sum of the distances between the consensus ranking and the ballots. Therefore the search for a consensus ranking needs 
to be based on a notion of distance between the rankings. There are several definitions of distance between rankings 
and many studies on the relations among them \cite{comparing}. Among these, the Kemeny distance $d_\text{Kem}$ \cite{kemeny} is widely used due to its robust properties \cite{hda}. Intuitively, when restricted 
to rankings without ties, $d_\text{Kem}$ is twice the minimum number of swaps of nearby candidates required to 
transform one ranking into another. Alternatively, it counts the number or pairwise preferences that do not match
in the two rankings. When ties appear, these count $\frac{1}{2}$ in the distance, if they do not match between the two rankings. 
Since our theory is rather insensitive to the type of distance used, we will conventionally use Kemeny distance to develop the discussion in the next sections (see the Supplementary Material for a discussion on other distances, see also \cite{truchon} and references therein).

The Condorcet consensus ranking $\mathbf{c}^*$ 
has been defined as the ranking (or more properly the ranking\emph{s}) minimizing the function
\begin{equation}\label{avg}
\mu(\mathbf{c}) = \frac{1}{n} \sum_{v=1}^n d_\text{Kem}(\mathbf{r}_v,\mathbf{c})\; ,
\end{equation}
in formulae, $\mathbf{c}^* = \text{argmin}\ \mu(\mathbf{c})$ (see \cite{monjardet}
for a review of mathematical methods in social choice theory).
The computation of $\mathbf{c}^*$ is in general a NP-hard problem, since the space of all possible 
rankings with ties grows faster than $m!$. In practice several polynomial time algorithms have been 
developed that return an approximated answer to the problem of selecting a consensus ranking. 
Most of these are the voting rules used in everyday applications. Among them it is worth recalling the
Pairwise comparison (or Copeland), Schulze and Tideman methods, which are 
perhaps the most used single-round ranked-ballot winner selection methods (they are all described in the Supplementary Material).

None of the above voting methods is perfectly fair (in the sense of Arrow's theorem), however they all return 
a ``reasonable'' consensus ranking, and this is why they are used in practical applications. 
Nonetheless some problems and inconsistencies remain unsolved: (i) different voting methods return different 
consensus rankings (this is the well known problem that the outcome of an election may very well depend on the electoral system); (ii) by returning a \emph{unique} consensus ranking, a lot of information about voter preferences is lost; (iii) often there are consensus rankings with a value of $\mu(\mathbf{c})$ very close to the optimal $\mu(\mathbf{c}^*)$, and it is unclear why they should be discarded.
It is worth noting that, in an election/survey with $n$ voters, fluctuations of $O(1/\sqrt{n})$ in $\mu(\mathbf{c})$ are somehow unavoidable: if $\mu(\mathbf{c}_1) < \mu(\mathbf{c}_2)$, but with $\mu(\mathbf{c}_2)-\mu(\mathbf{c}_1) \sim 1/\sqrt{n}$, then choosing $\mathbf{c}_1$ as the consensus ranking instead of $\mathbf{c}_2$ is equivalent to taking a decision based on the toss of a coin.

\paragraph*{A new dimension for choosing the consensus ranking}
In order to solve the above problems we suggest to consider as valid consensus rankings all the 
rankings $\mathbf{c}$ close enough to the optimal one (i.e., those for which $\mu(\mathbf{c})-\mu(\mathbf{c}^*) \sim 1/\sqrt{n}$), 
and we introduce a new \emph{dimension} to select the best among these valid consensus rankings.
Our idea is that not only the global (i.e., societal) number of satisfied preferences is to be maximized, but also 
each individual voter should have more or less the same number of satisfied preferences.
With this aim we propose to consider also the voter-to-voter satisfaction variability (standard deviation) as
\begin{equation}\label{std}
\sigma(\mathbf{c}) = \sqrt{\frac{1}{n} \sum_{v=1}^n \big[d_\text{Kem}(\mathbf{r}_v,\mathbf{c}) - \mu(\mathbf{c})\big]^2} = \sqrt{\frac{1}{n} \sum_{v=1}^n d_\text{Kem}(\mathbf{r}_v,\mathbf{c})^2 - \mu(\mathbf{c})^2}\;.
\end{equation}
If $\sigma(\mathbf{c})=0$, the consensus ranking $\mathbf{c}$ satisfies equally each voter; while, if $\sigma(\mathbf{c})$ is large, then 
there are voters more satisfied and others less satisfied than the average. Clearly the smaller is $\sigma(\mathbf{c})$ 
the more egalitarian is $\mathbf{c}$.


%
\begin{figure}
\centering
\includegraphics[width=0.47\textwidth]{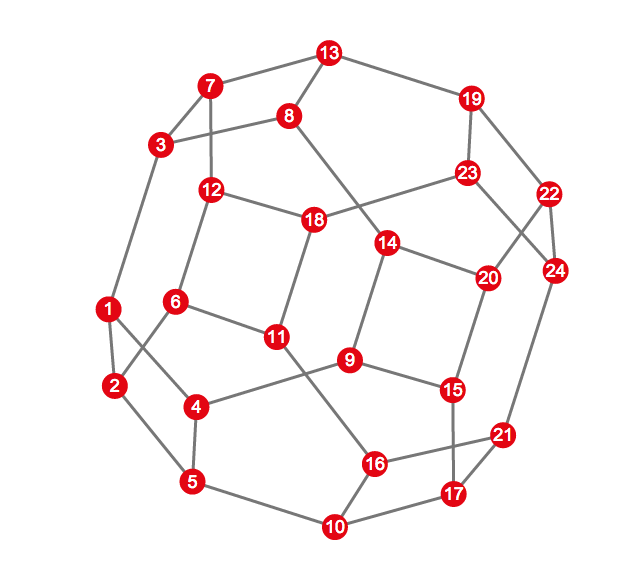}
\includegraphics[width=0.47\textwidth]{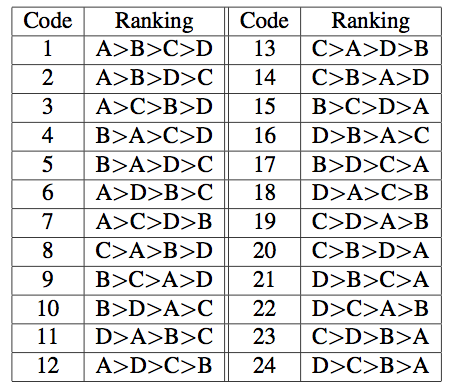}
\\
\includegraphics[width=0.47\textwidth]{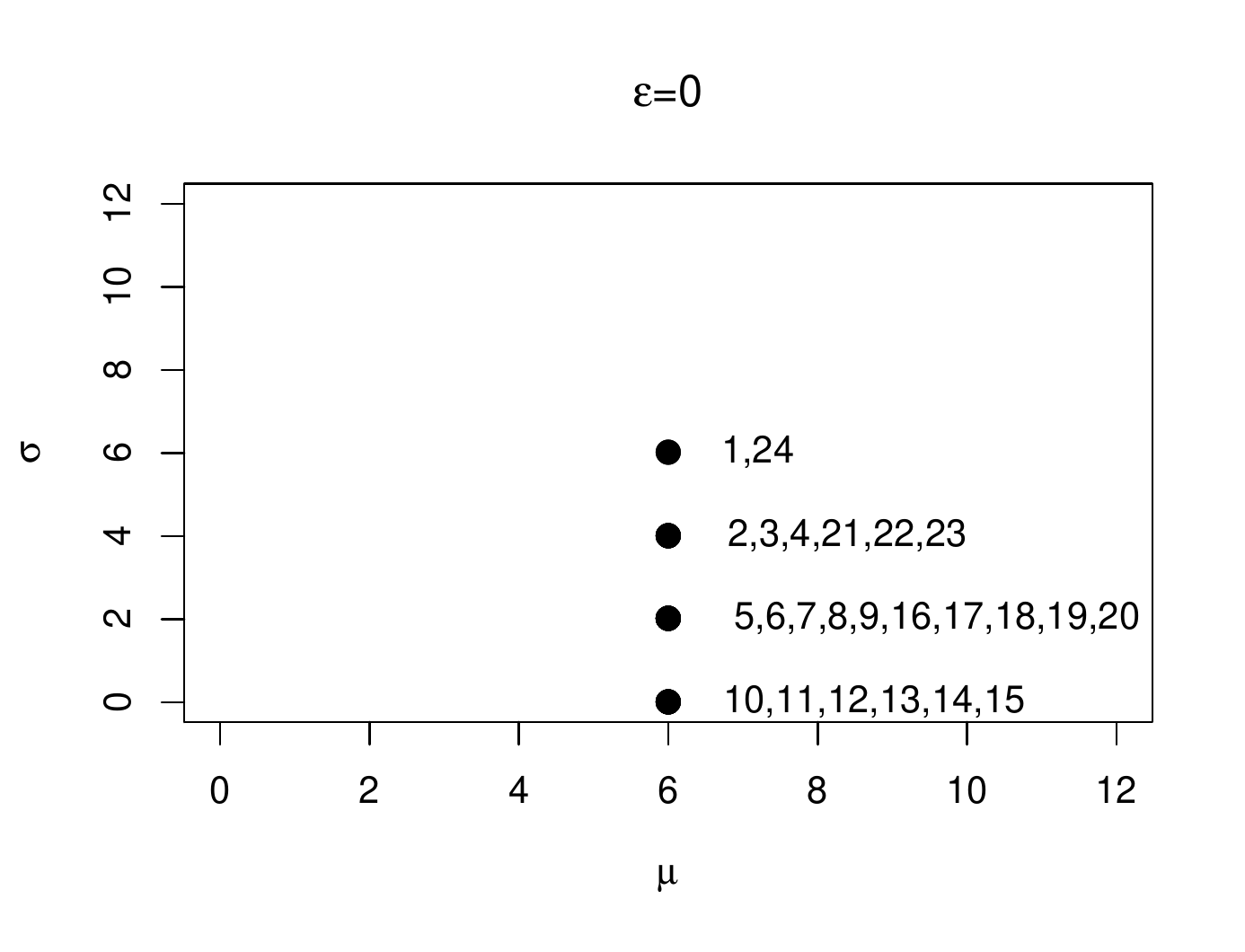}
\\
\includegraphics[width=0.47\textwidth]{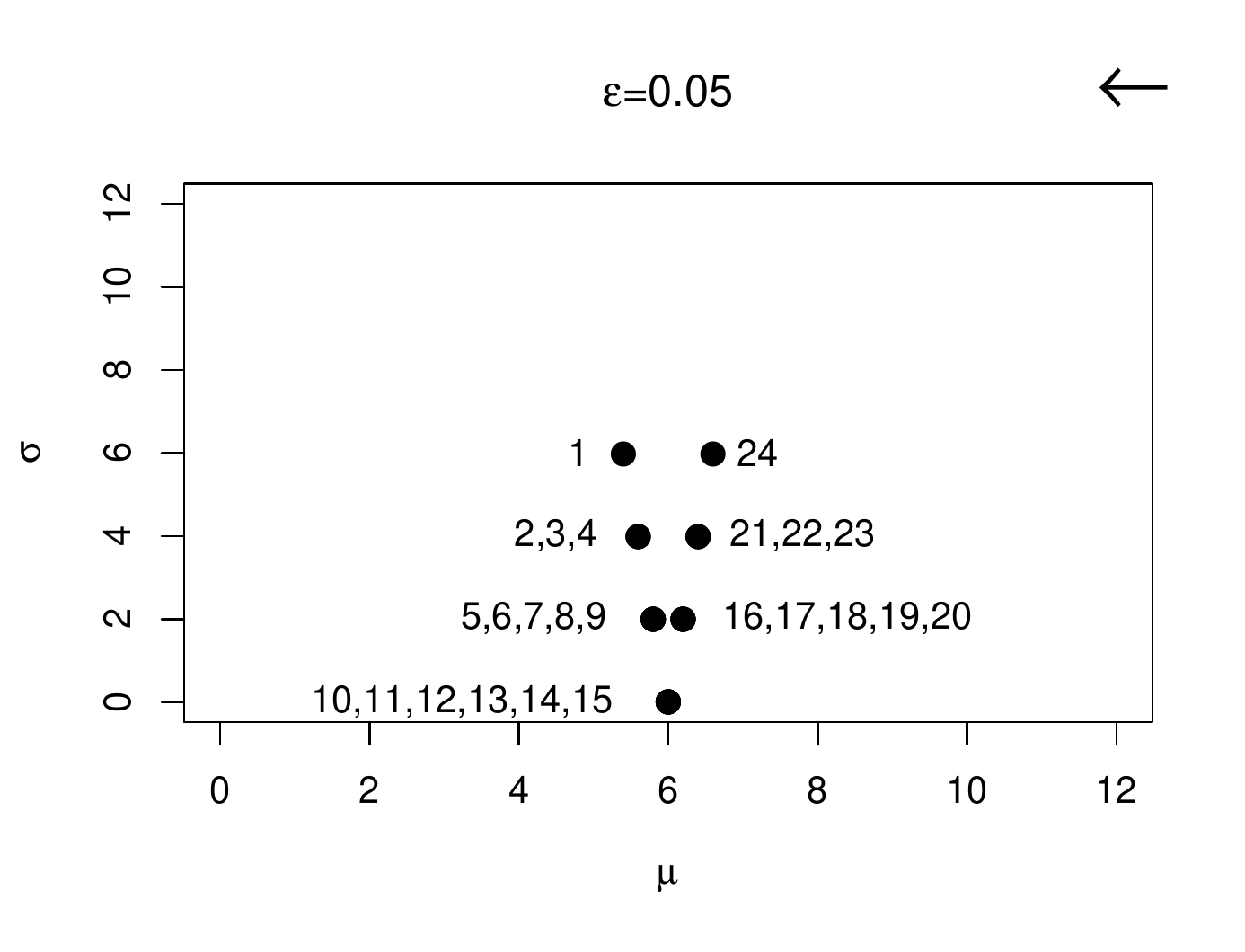}
\includegraphics[width=0.47\textwidth]{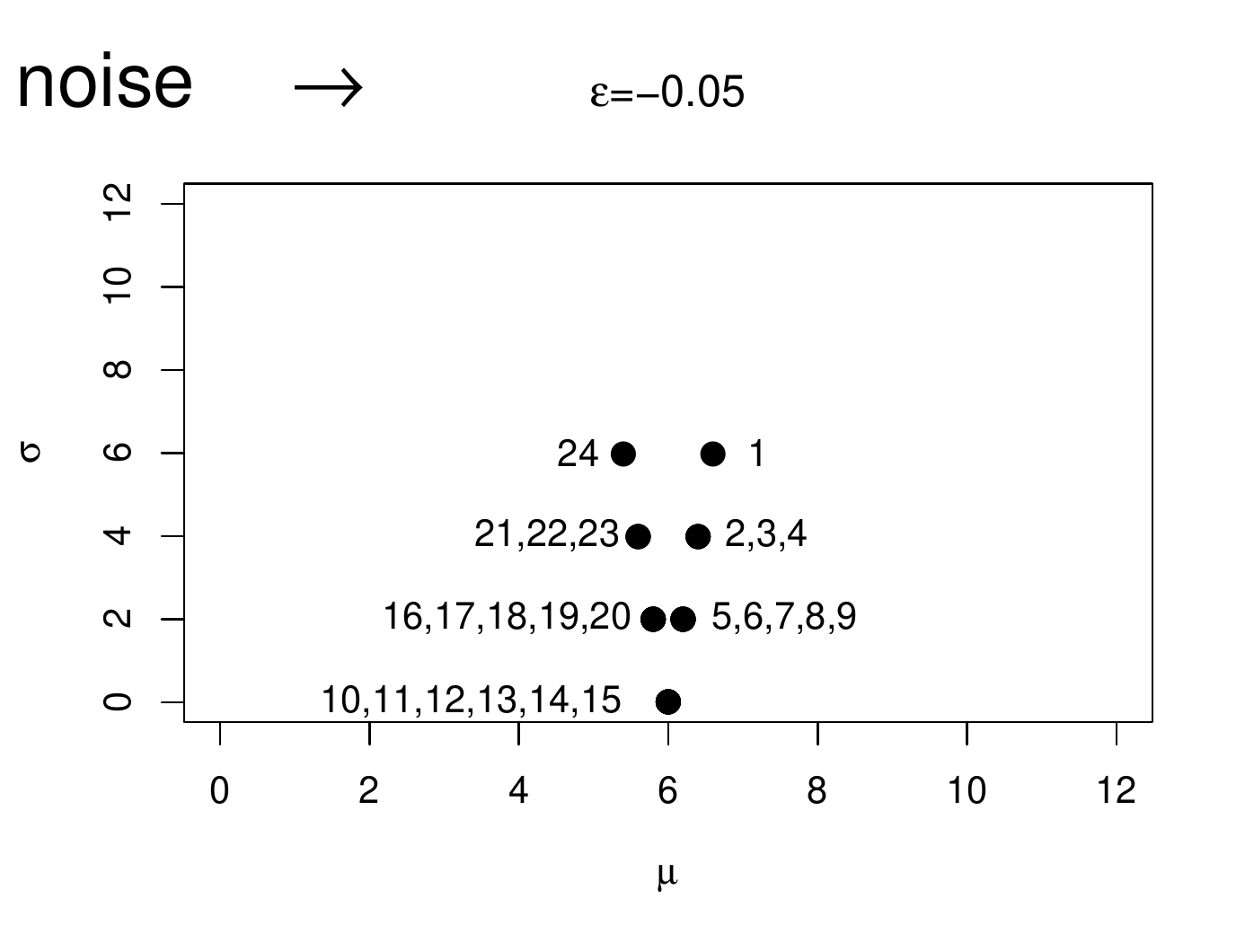}
\caption{A very simple example with four candidates. There are 24 possible rankings without ties (listed in panel (b)), whose relative distances are given by the graph in panel (a). In panel (c) we report $(\mu,\sigma)$ values for possible consensus rankings in case the electorate is equally polarized on opposite ballots (codes 1 and 24); panels (d) and (e) have been computed by adding a small noise to the perfectly balanced situation.  Our web platform allows one to interact with panels \href{http://www.sapienzaapps.it/rateit.php?ex=1c}{\underline{c}},  \href{http://www.sapienzaapps.it/rateit.php?ex=1d}{\underline{d}} and \href{http://www.sapienzaapps.it/rateit.php?ex=1e}{\underline{e}}.}
\label{fourCand}
\end{figure}

To illustrate the new criterion, we start with a very simple (and almost paradoxical) example. We consider an election with $m=4$ candidates and we do not allow for ties; the number of possible rankings is $m!=24$, as shown in the table included in Figure \ref{fourCand}. The distance between these 24 rankings can be easily visualized in the same figure, top left panel, which includes a graph where each vertex corresponds to a ranking, with an edge connecting rankings at distance 2 (differing only by a swap of two neighbouring candidates).  For rankings at distance larger than 2, it is enough to count the edges along the shortest path between the rankings in this graph.

Suppose the electorate is equally polarized on two opposite rankings: half of the voters rank candidates $\mathsf{A>B>C>D}$ and the other half $\mathsf{D>C>B>A}$. 
A simple calculation shows that any possible ranking has $\mu(\mathbf{c})=6$, therefore there is no way to choose one of them according to the Condorcet criterion alone.
However the 24 possible consensus rankings have very different $\sigma(\mathbf{c})$ as can be seen in the middle panel of Figure~\ref{fourCand}:
the point with largest $\sigma(\mathbf{c})$ corresponds to rankings $\mathsf{A>B>C>D}$ and $\mathsf{D>C>B>A}$ that fully satisfy half of the voters and fully deceive the second half,
while the point with $\sigma(\mathbf{c})=0$ corresponds to the six rankings that are at the same distance from the ballots, thus satisfy them equally well. 
It is clear that the latter are the more egalitarian consensus rankings.
In other words, spreading satisfaction as equally as possible among voters,
i.e.\ minimizing $\sigma(\mathbf{c})$, is a new criterion to select the consensus ranking that deserves, at least, the same consideration as the Condorcet criterion of minimizing $\mu(\mathbf{c})$. 

Even more interesting is the case when some noise is added to the example above. For instance we can consider small fluctuations in the number of electors participating to the poll, resulting in a fraction $\frac{1}{2}+\epsilon$ of voters ranking the candidates as $\mathsf{A>B>C>D}$ and the complement fraction $\frac{1}{2}-\epsilon$ ranking them as $\mathsf{D>C>B>A}$.
For an election with $n$ voters a noise of order $\epsilon=O(1/\sqrt{n})$ is somehow unavoidable.
In lower panels in Figure~\ref{fourCand} we report $\mu(\mathbf{c})$ and $\sigma(\mathbf{c})$ values for the 24 possible consensus rankings. For $\epsilon>0$, the small unbalance decreases $\mu$ for ranking $\mathsf{A>B>C>D}$, making it the consensus ranking under the Condorcet criterion. For $\epsilon<0$, the opposite ranking would win. The difference between the two cases is, however, only due to noise; so selecting a consensus ranking by strictly minimizing $\mu(\mathbf{c})$ would be equivalent to selecting the winner on a coin toss.  Rankings with lower $\sigma(\mathbf{c})$, as lower panels in Figure \ref{fourCand} shows, are much less sensitive to noise: by minimizing $\sigma(\mathbf{c})$ one gets always the same consensus rankings independently on the noise.
This is a very important observation in favour of the new criterion, given that a fair voting system should be robust to noise induced fluctuations.

Although very simplified, the example above contains in a stylized form the relevant facts we have observed in real data, to be discussed below.

\paragraph*{Analysis of data from real polls}
We have two criteria for the identification of the best consensus ranking: minimizing $\mu(\mathbf{c})$ and minimizing $\sigma(\mathbf{c})$ 
(among rankings of small $\mu(\mathbf{c})$).
In general is not possible to find a consensus ranking satisfying both criteria, and some compromise must be adopted, as we will exemplify with data from real polls.

\begin{figure}
\centering
\fbox{\includegraphics[width=0.7\textwidth]{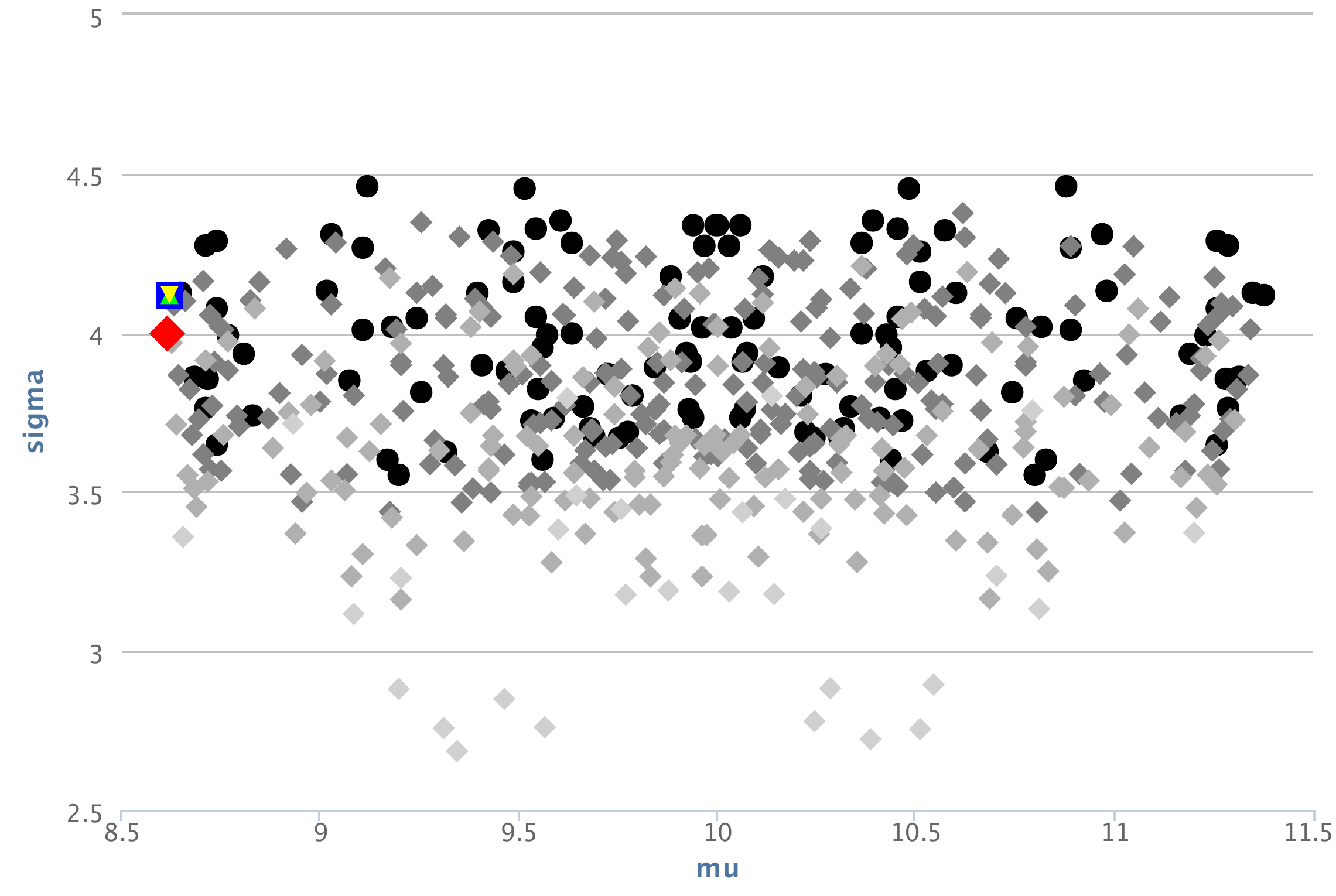}}
\fbox{\includegraphics[width=0.7\textwidth]{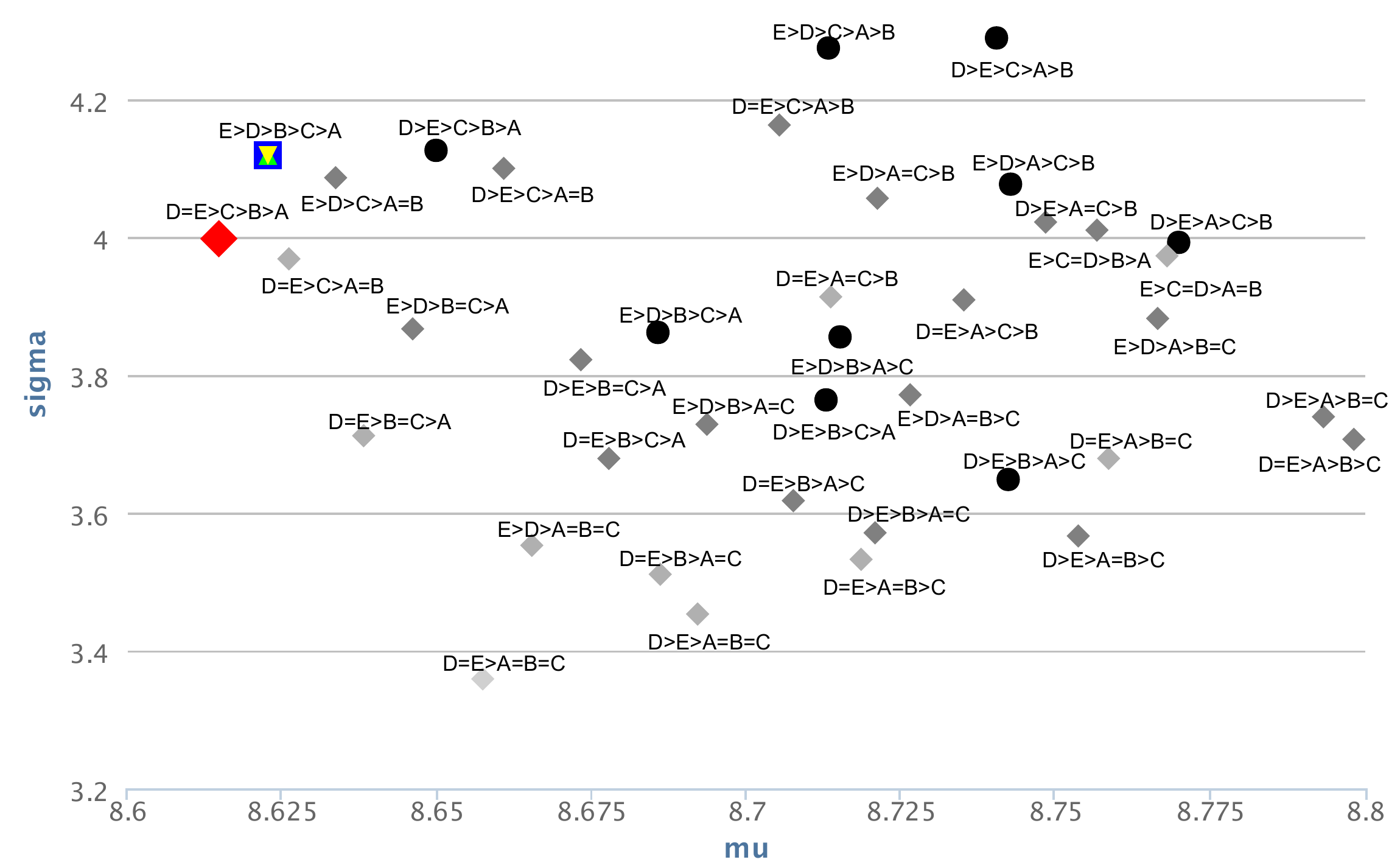}}
\\
\hspace{1cm}{\large$\bullet$} Rankings with no ties \hfill
{\color{red}\large$\vardiamond$} Condorcet solution ($\mathsf{D\!=\!E\!>\!C\!>\!B\!>\!A}$)\\
\hspace{1cm}{\color{grey1}$\vardiamond$} Rankings with one tie  \hfill
{\color{blue}$\blacksquare$} Copeland solution ($\mathsf{E\!>\!D\!>\!C\!>\!B\!>\!A}$)\\
\hspace{1cm}{\color{grey2}$\vardiamond$} Rankings with two ties \hfill
{\color{yellow}$\blacktriangledown$} Schulze solution ($\mathsf{E\!>\!D\!>\!C\!>\!B\!>\!A}$)\\
\hspace{1cm}{\color{grey3}$\vardiamond$} Rankings with three ties \hfill
{\color{green}\hspace{1cm}$\blacktriangle$} Tideman solution ($\mathsf{E\!>\!D\!>\!C\!>\!B\!>\!A}$)\\
\hspace{1cm}{\color{grey4}$\vardiamond$} Rankings with four ties \hfill \, \\
\caption{Aggregation of 24921 ballots rankings 5 jokes. The upper panel shows the entire set of possible rankings, except for $\mathsf{A\!=\!B\!=\!C\!=\!D\!=\!E}$, at position (9.81,0.63) which is omitted for visualization purposes.  The lower panel zooms in the leftmost part of the first plot. Solutions found by Copeland, Tideman and Schulze coincide in this example. None of these is optimal under the Condorcet criterion  of minimizing $\mu$ or under the new criterion of minimizing $\sigma$. \href{http://www.sapienzaapps.it/rateit.php?ex=2a}{\underline{Here}} you may interact with this figure.}
\label{joke}
\end{figure}

The first dataset consists of ratings for jokes from the Jester database\cite{jester}. The full dataset is made of 100 jokes rated by 24938 users. Ratings are continuous values between -10 and 10. We have selected the five jokes rated by most users, and considered only those users who rated all five jokes, resulting in 24921 voters. For each voter, the ballot is obtained by ranking the 5 jokes according to the continuous-valued rating.

In the upper panel of Figure~\ref{joke} we show the $\big(\mu(\mathbf{c}),\sigma(\mathbf{c})\big)$ values for all possible consensus rankings of the $m=5$ jokes: the 120 black circles correspond to rankings without ties, while gray diamonds are the 421 rankings with ties.
One ranking was excluded from the plot, for better visualization: ranking $\mathsf{A\!=\!B\!=\!C\!=\!D\!=\!E}$ at position (9.81,0.63). 
The consensus ranking minimizing $\mu(\mathbf{c})$ is $\mathbf{c}^*:\mathsf{D\!=\!E\!>\!C\!>\!B\!>\!A}$ and has $\mu(\mathbf{c}^*)=8.615$. However, close to $\mathbf{c}^*$ we see a cloud of points with small values of $\mu(\mathbf{c})$. The lower panel in Figure~\ref{joke} zooms over this set of rankings, all having a distance from the Condorcet optimum $\mathbf{c}^*$, comparable with $O(1/\sqrt{n})$ fluctuations. So, from the point of view of the Condorcet criterion, all these rankings are equally good within the noise.
On the contrary they show a much larger variation in $\sigma(\mathbf{c})$, that changes between 3.36 and 4.29, allowing for a better consensus ranking selection by minimizing $\sigma(\mathbf{c})$.
The consensus ranking minimizing $\sigma(\mathbf{c})$ in this region is $\mathsf{D\!=\!E\!>\!A\!=\!B\!=\!C}$ with coordinates $(\mu,\sigma)=(8.66,3.36)$. It seems to convey all the relevant information contained in this set of low $\mu(\mathbf{c})$ rankings: indeed the only information shared by all the rankings in the lower panel of Figure~\ref{joke} is that jokes $\mathsf{D}$ and $\mathsf{E}$ are better than jokes $\mathsf{A}$, $\mathsf{B}$ and $\mathsf{C}$.
Any consensus ranking more refined that $\mathsf{D\!=\!E\!>\!A\!=\!B\!=\!C}$ would just amplify the noise, rather than providing further useful information.

Three commonly used winner selection methods were also applied to the data (Copeland, Schulze and Tideman\footnote{These three algorithms have been chosen also because they never rank first a Condorcet loser.}), and the corresponding consensus rankings are marked in Figure~\ref{joke}. All of them rank jokes as $\mathsf{E\!>\!D\!>\!C\!>\!B\!>\!A}$ with $(\mu,\sigma)=(8.62,4.12)$. This consensus ranking differs from $\mathbf{c}^*$, the Condorcet consensus ranking, and it has a quite large $\sigma(\mathbf{c})$ value, hence being among the less egalitarian rankings.

In applying the criterion of minimizing $\sigma(\mathbf{c})$ one has to be careful, because this criterion tends to select consensus rankings with ties (gray diamonds are on average below black circles in Figure~\ref{joke}). If ties are not allowed in the consensus ranking, one should focus only on black points in Figure~\ref{joke}: even in this case, the consensus ranking $\mathsf{D\!>\!E\!>\!B\!>\!A\!>\!C}$ with $(\mu,\sigma)=(8.74,3.65)$ looks much more egalitarian than the consensus ranking $\mathsf{E\!>\!D\!>\!C\!>\!B\!>\!A}$ found by common voting methods: it gains more than 10\% in $\sigma(\mathbf{c})$, while loosing just 1\% in $\mu(\mathbf{c})$.
The final decision on which rule is to be used to select the consensus ranking is left to the organizers of the poll/survey, but clearly a plot in the $(\mu,\sigma)$ plane is much more informative than any previously available method.

\begin{figure}[t]
\centering
\includegraphics[width=0.8\textwidth]{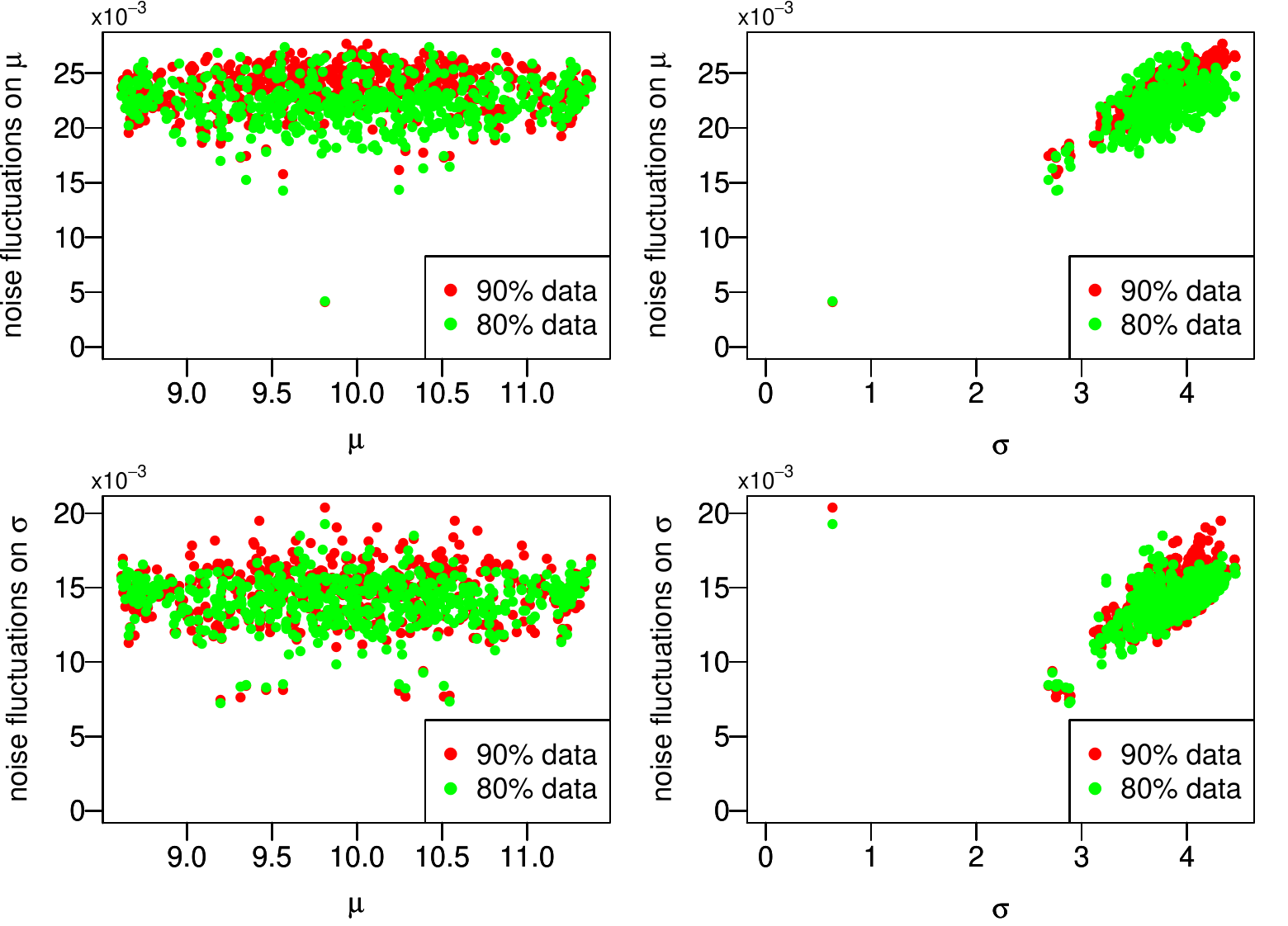}
\caption{Uncertainties on the values of $\mu(\mathbf{c})$ and $\sigma(\mathbf{c})$ as obtained from resampling experiments using either 80\% or 90\% of the original data ($\alpha=0.2$ and $\alpha=0.1$ respectively). Rankings with smaller $\sigma(\mathbf{c})$ are more reliable, since they have smaller fluctuations.}
\label{stab}
\end{figure}

Similar to the simple example discussed earlier, the data from real polls also show that consensus rankings of smaller $\sigma(\mathbf{c})$ are less sensitive to noise. In this case we investigate the effect of small fluctuations in participation by using a subsampling procedure: from the joke ratings provided by 24921 users, a fraction $\alpha$ of randomly chosen votes has been removed, and $\mu(\mathbf{c})$ and $\sigma(\mathbf{c})$ recomputed. Resampling was repeated 100 times with $\alpha=0.1$ and $\alpha=0.2$. From the variations of $\mu$ and $\sigma$ between different subsamplings we may compute the noise fluctuations on $\mu$ and $\sigma$ (see Supplementary Material for a detailed definition of the fluctuation scaling). In Figure \ref{stab} these fluctuations are reported, showing a very clear and strong correlation with the value of $\sigma(\mathbf{c})$.
A good consensus ranking should be as robust as possible to noise produced by fluctuations in e.g.\ the number of voters. For example, suppose a poll/survey is run for 10 days, then the outcome of the survey is reliable if it does not change sensibly in case the data were collected for one or two days less.
What we observe in Figure~\ref{stab} is that noise sensitivity is larger for points of large $\sigma(\mathbf{c})$, while no relation can be observed between noise sensitivity and $\mu$. So, choosing a consensus ranking according to the new criterion of minimizing $\sigma(\mathbf{c})$, provides in general a result much more robust to noise (e.g. unavoidable fluctuations in the number of participants to the poll/survey/election). We also analysed noise in opinion for this dataset, similar to the analytical example of Figure \ref{fourCand}, and results show same robustness for rankings with lower $\sigma(\mathbf{c})$ (see Supplementary Material for details).

\begin{figure}
\centering
\fbox{\includegraphics[width=0.7\textwidth]{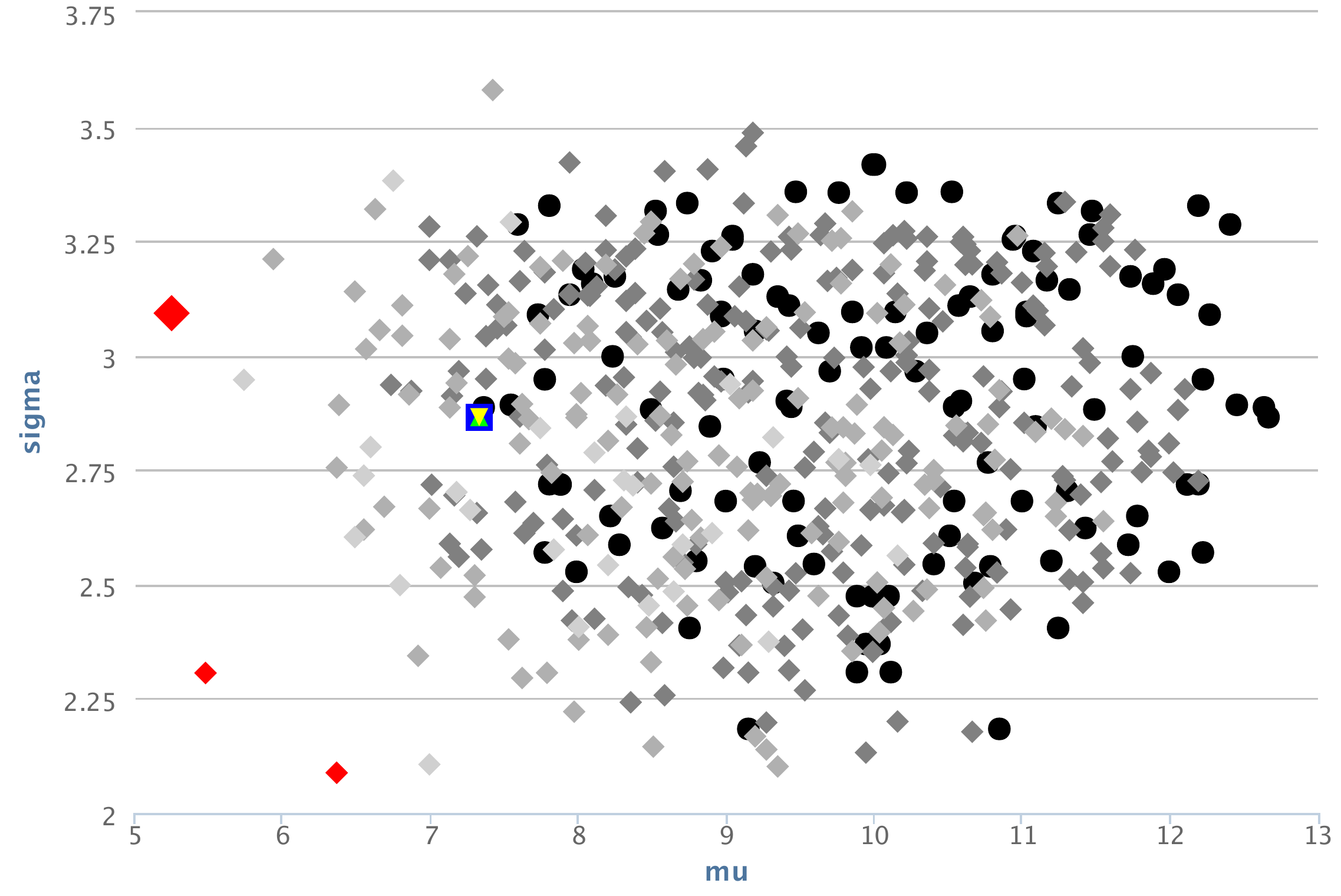}}
\\
\hspace{1cm}{\large$\bullet$} Rankings with no ties \hfill
{\color{red}\large$\vardiamond$} Condorcet solution ($\mathsf{A\!=\!B\!=\!C\!=\!D\!>\!E}$)\\
\hspace{1cm}{\color{grey1}$\vardiamond$} Rankings with one tie  \hfill
{\color{blue}$\blacksquare$} Copeland solution ($\mathsf{B\!>\!C\!>\!A\!>\!D\!>\!E}$)\\
\hspace{1cm}{\color{grey2}$\vardiamond$} Rankings with two ties \hfill
{\color{yellow}$\blacktriangledown$} Schulze solution  ($\mathsf{B\!>\!C\!>\!A\!>\!D\!>\!E}$)\\
\hspace{1cm}{\color{grey3}$\vardiamond$} Rankings with three ties \hfill
{\color{green}\hspace{1cm}$\blacktriangle$} Tideman solution ($\mathsf{B\!>\!C\!>\!A\!>\!D\!>\!E}$)\\
\hspace{1cm}{\color{grey4}$\vardiamond$} Rankings with four ties \hfill 
{\color{red}\footnotesize$\vardiamond$ }Optimal set\\
\caption{Aggregation of 930 ballots ranking 5 movies. Here rankings with ties play an important role. Again the solution found by standard voting methods (Copeland, Schulze, Tideman) is far from the optimal set. \href{http://www.sapienzaapps.it/rateit.php?ex=4}{\underline{Here}} you may interact with this figure.}
\label{movie}
\end{figure}

The second example from real polls considers the rankings of 5 movies provided by 930 users. These are a subset of a larger database consisting of 1,000,209 ratings from 6040 users for 3952 movies \cite{movies}. Here, users rated the movies on a discrete scale from 1 to 5.  As before, we sorted the movies for each user, to obtain the ballots. Since equal ratings are very probable here, given that only 5 possible rating values exist, many ballots have ties. 

Once again the plot in $(\mu,\sigma)$, shown in Figure~\ref{movie} is very informative. First of all we notice that consensus rankings with ties, although having much smaller values of $\mu(\mathbf{c})$, are not chosen by any commonly used voting method. Moreover the optimal consensus ranking according to the Condorcet criterion seems to have a very large value for $\sigma(\mathbf{c})$. There are a few other rankings worth to be considered, that have slightly higher $\mu(\mathbf{c})$ but much lower $\sigma(\mathbf{c})$. Indeed we identify a \emph{set of optimal rankings} (red diamonds in Figure~\ref{movie}) combining the two criteria. These optimal rankings start from the Condorcet ranking $\mathbf{c}^*$ and include other rankings in the bottom left part of the plot, that cannot be improved in terms of both $\mu(\mathbf{c})$ and $\sigma(\mathbf{c})$ (the Supplementary Material includes a more formal definition of this sequence). In the example from the movie data, two additional rankings should be considered, along with the Condorcet ranking, to be part of the optimal set: these are $\mathsf{A\!=\!B\!=\!C\!=\!D\!=\!E}$ with $(\mu,\sigma)=(5.48,2.30)$, and $\mathsf{B\!>\!A\!=\!C\!=\!D\!=\!E}$ with $(\mu,\sigma)=(6.37,2.08)$, both are red-marked in the bottom left corner of the plot. We suggest that the consensus ranking should be selected from the optimal set, and the choice should be made after careful analysis of the $(\mu,\sigma)$ plot.
The set of optimal rankings resembles somehow the Pareto efficient frontier used in economic theory \cite{Pareto}.

\begin{figure}[t]
\centering
\fbox{\includegraphics[width=0.7\textwidth]{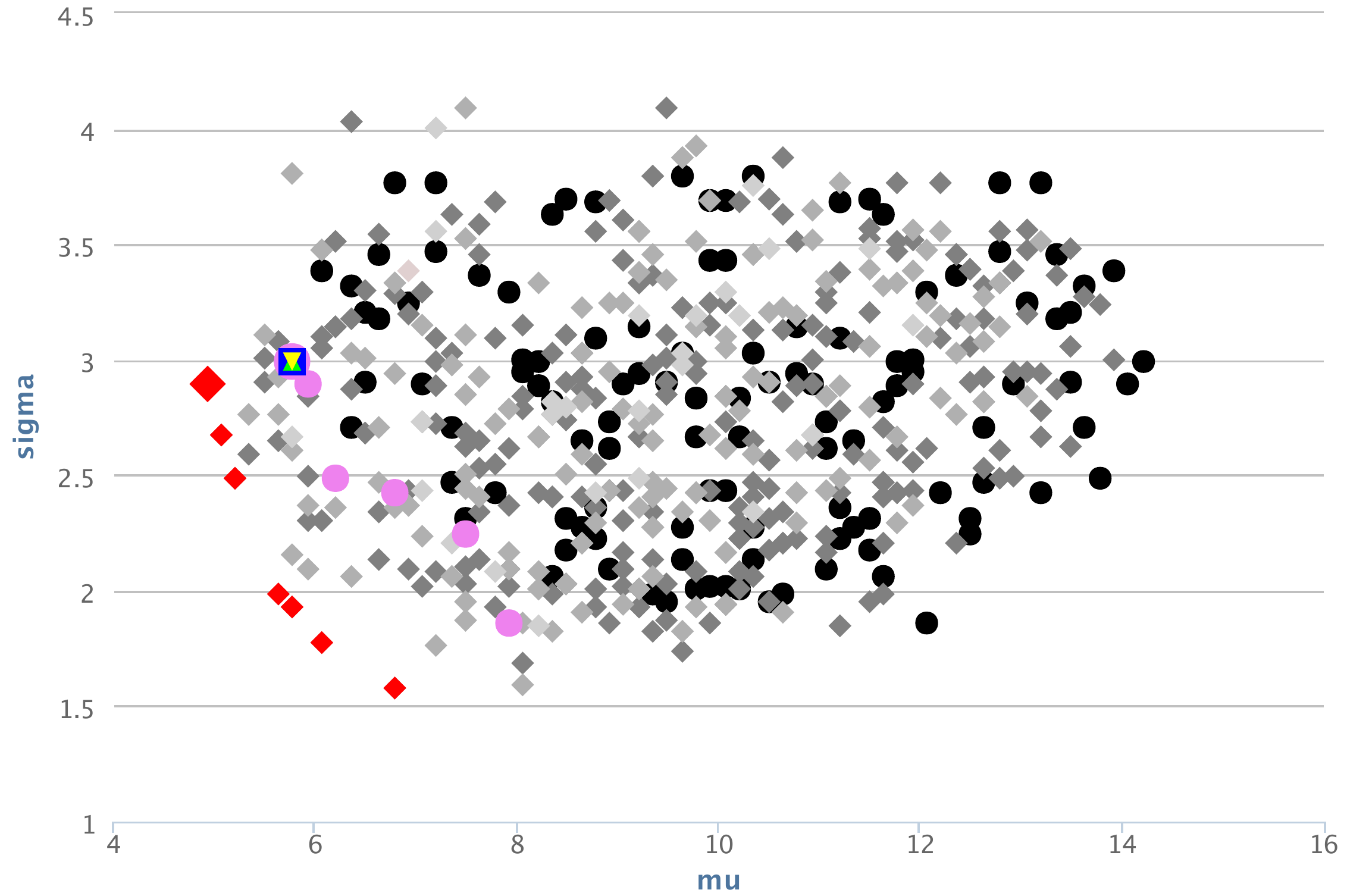}}
\\
\hspace{1cm}{\large$\bullet$}  Rankings with no ties \hfill
{\color{red}\large$\vardiamond$} Condorcet solution ($\mathsf{C\!=\!D\!=\!E\!>\!A\!=\!B}$)\\
\hspace{1cm}{\color{grey1}$\vardiamond$} Rankings with one tie  \hfill
{\color{blue}$\blacksquare$} Copeland solution ($\mathsf{C\!>\!E\!>\!D\!>\!B\!>\!A}$)\\
\hspace{1cm}{\color{grey2}$\vardiamond$} Rankings with two ties \hfill
{\color{yellow}$\blacktriangledown$} Schulze solution  ($\mathsf{C\!>\!E\!>\!D\!>\!B\!>\!A}$)\\
\hspace{1cm}{\color{grey3}$\vardiamond$} Rankings with three ties \hfill
{\color{green}\hspace{1cm}$\blacktriangle$} Tideman solution ($\mathsf{C\!>\!E\!>\!D\!>\!B\!>\!A}$)\\
\hspace{1cm}{\color{grey4}$\vardiamond$} Rankings with four ties \hfill 
{\color{red}\footnotesize$\vardiamond$ }Optimal set {\color{violet}\large$\bullet$ }Optimal without ties\\
\caption{AIRESIS data. \href{http://www.sapienzaapps.it/rateit.php?ex=5}{\underline{Here}} you may interact with this figure.}\label{airesis}
\end{figure}

Both examples above have a large number of voters and one may think that the complex behaviour we have illustrated could be due to the large number of voters.
This is not actually the case, as we are going to show with an example from a poll with a small number of voters ($n=14$), that ranked $m=5$ alternatives.
This is a poll organized on the Airesis platform \cite{airesis}, which is a web platform freely available to organizations to manage internal decision making.
The data shown in Figure~\ref{airesis} represent a real election where the consensus ranking has been selected according to the Schulze method.
The first evidence is that the consensus ranking of that election (Schulze) is far from the optimal one: the Condorcet optimal consensus ranking is better (i.e.\ lower) both in $\mu(\mathbf{c})$ and $\sigma(\mathbf{c})$. Additionally, a large number of rankings are part of the optimal set, defined previously, and marked with red diamonds in the plot, which should be taken into consideration.
Even willing to restrict to consensus rankings without ties (this is an election, and ties may be problematic for the decision process), it is clear that the consensus ranking selected by Schulze, $\mathsf{C\!>\!E\!>\!D\!>\!B\!>\!A}$ with $(\mu,\sigma)=(5.78,2.99)$, has a quite large $\sigma(\mathbf{c})$ with respect to consensus rankings $\mathsf{C\!>\!D\!>\!E\!>\!B\!>\!A}$ with $(\mu,\sigma)=(5.92,2.89)$, and $\mathsf{C\!>\!E\!>\!D\!>\!A\!>\!B}$ with $(\mu,\sigma)=(6.21,2.48)$. The latter correspond to the two leftmost purple circles in Figure~\ref{airesis}.

\paragraph*{Discussion}
We have proposed to analyze voting results by plotting potentially winning rankings 
on the plane $(\mu,\sigma)$ in such a way that both the Condorcet criterion and the new criterion that we have introduced can be considered at the same time in order to identify the optimal consensus ranking.
To help in this new analysis we have set up a webpage with an interactive tool that produces the graph in the $(\mu,\sigma)$ plane \cite{site}, once the list of ranked ballots is given as input.\footnote{All plots in this manuscript, using the standard Kemeny distance, are based on those produced by the web tool.} A publicly available Android mobile application has also been developed, to facilitate organisation of large scale ranked-ballot polls and collection of new data for future studies \cite{rankit}.
We have analyzed many different datasets coming from real polls and in general the plots in the $(\mu,\sigma)$ plane are similar to those shown above. Moreover we expect polynomial time algorithms can be developed that minimize (approximately) both $\mu(\mathbf{c})$ and $\sigma(\mathbf{c})$ in analogy to presently used voting rules that tend to minimize only $\mu(\mathbf{c})$.

Once the graph in the $(\mu,\sigma)$ plane is available, we believe any good consensus ranking should be chosen from the {\it optimal set}. A point belongs to the optimal set if no other point exists improving both in $\mu(\mathbf{c})$ and $\sigma(\mathbf{c})$ or improving only one of them while keeping the other constant. 
This set has been red-marked in the examples above and it extends from the Condorcet optimal ranking $\mathbf{c}^*$, that minimizes $\mu(\mathbf{c})$, 
to the ranking $\mathbf{\overline c}$ minimizing $\sigma(\mathbf{c})$.
The meaning of moving along this set should be clear: while the ranking $\mathbf{c}^*$ maximizes total societal satisfaction 
ignoring individual satisfaction, $\mathbf{\overline c}$ is the more egalitarian in terms of individual satisfaction
regardless of the total satisfaction. There are polls, like political elections, where the consensus ranking must produce a unique winner 
among the candidates. In this case one can restrict the analysis to rankings having no tie at the first position and a line of optimal rankings can be defined as well in this subset of rankings. 

It is important to stress that we are not claiming that $\mathbf{\overline c}$ should be the consensus ranking:  
often $\mu(\mathbf{\overline c})$ is much larger that $\mu(\mathbf{c}^*)$ and the optimal consensus ranking is actually in the middle of the optimal set.
Instead, we are proposing a new tool that provides a quantitative meaning to each possible choice. Which consensus ranking 
should be chosen among the optimal set is no longer a technical matter, it is rather a decision to be taken by the
people in charge and the criteria may change according to the domains: political elections, marketing, web page ranking, etc.
In some cases, like for instance in political election, the decision on which point to select along this line must be taken \emph{before} 
the poll is run. 

The cases where the plot in the $(\mu,\sigma)$ plane is even more useful is when the final decision can be taken \emph{after} the poll/survey is run. In this case having a data aggregation like the one we are presenting in terms of $\mu(\mathbf{c})$ and $\sigma(\mathbf{c})$ provides a lot of information and allows for a much better choice.
A typical example is when politicians want to decide a list of priorities based on suggestions coming from the electorate: the politicians can run a poll/survey among the electorate and this would determine the optimal rankings, leaving to the politicians the final choice of the consensus ranking, to be chosen among those. We believe this is an ideal compromise between taking in serious consideration the desiderata of the electorate (the line of optimal consensus rankings is fully determined by the votes) and leaving the political decision to those in charge.

It is worth mentioning that the applications where technical tools provide a set of optimal preferences among which the final choice is left to the user 
are not new in other fields. For example in quantitative financial risk management the mathematical analysis produces a risk-return curve 
(called efficient frontier \cite{mar}) and the choice of a point along such a curve is left to the investor. From a different perspective a 
voting theory purely based on the maximization of voters satisfactions would be equivalent, in political economy, to the maximization of 
total wealth in a country regardless of its distribution and welfare criteria. 

The voting method we have presented here provides 
an efficient technical tool to determine the line of optimal rankings, among which a political decision has to be taken.
While it is generally understood and acknowledged that democratic organizations should not only maximize their 
goods but also distribute them as equally as possible, such awareness did not lead so far to a proper solution 
in social choice theory. We believe therefore that the quantitative method we have introduced is a fundamental tool 
to apply democratic principles, especially in voting processes.  

\paragraph*{Acknowledgments} We thank Flavio Chierichetti for drawing our attention to the rank aggregation problem, and the Airesis platform for providing access to their data. This work has received financial support from the Italian Research Ministry through the FIRB projects 
No. RBFR086NN1 and RBFR10N90W and PRIN project No. 2010HXAW77. Mobile application development was performed by Federico Ponzi, with partial financial support from New York University Shanghai.

\newpage


\begin{center}
\section*{Supplementary Material for\\ \vskip .3truecm
\textit{Egalitarianism in the rank aggregation problem:\\ a new dimension for democracy}}
\end{center}
\vspace{0.5cm}

\subsection*{Combinatorics}

The set of rankings without ties for $m$ candidates, ${\cal R}_m$,  has cardinality $m!$. 
Let us call $T_m$ the cardinality of the set of rankings including ties, $\overline{\cal R}_m$. $T_m$ are sometimes called Fubini, or Cayley numbers. 
One can show (see \cite{oeis} and references therein) that their exponential generating function is
$$
F(x) \; = \; \sum_{m=0}^{\infty}\frac{T_m}{m!}x^m=\frac{1}{2-e^x} \; ,
$$
whose radius of convergence is $\ln 2$. This can be used to find $T_m$ from derivatives and gives
$T_0=1$, $T_1=1$, $T_2=3$, $T_3=13$, $T_4=75$, $T_5=541$, $T_6=4683$, $T_7=47293$, $T_8=545835$, $T_9=7087261$, 
$T_{10}=102247563$, etc. These numbers grow according to the formula
$$
T_m \; \simeq \; \frac{m!}{2(\ln 2)^{m+1}} \; \simeq \; (1.44)^m m! \; ,
$$
with a sub-leading correction decaying exponentially fast
$$
\left(T_m - \frac{m!}{2(\ln 2)^{m+1}}\right) \frac{1}{T_m} \simeq (0.11)^m\; .
$$
One can also consider the set of rankings with $l$ ties, where $0\le l \le m-1$: $\overline{\cal R}_m^{(l)}$. Clearly ${\cal R}_m=\overline{\cal R}_m^{(0)}$
and $\overline{\cal R}_m=\cup_{l=0}^{m-1}\overline{\cal R}_m^{(l)}$. Another interesting set for applications is the set $\widehat{\cal R}_m$ containing rankings where the first candidate is untied. For each set of rankings our method provides a subset of optimal rankings according to the following definition.

\subsection*{Optimal set}

For two rankings $\mathbf{s}$ and $\mathbf{r}$ in ${\cal S}$, we say $\mathbf{s}$ \emph{improves} $\mathbf{r}$ if $\sigma(\mathbf{s})\leq\sigma(\mathbf{r})$ and $\mu(\mathbf{s})\leq\mu(\mathbf{r})$, and at least one of the two inequalities is strict. The optimal set ${\cal O}_{\cal S}$ is the set of points in ${\cal S}$ that cannot be improved by other elements of ${\cal S}$.

In the web platform that we have developed \cite{site2} we show the global optimal set ${\cal O}_{\overline{\cal R}_m}$ (red diamonds)  and the one with no ties ${\cal O}_{{\cal R}_m}$ (purple circles).
In other contexts, like engineering or economics, the optimal set of vectors of a $d$-dimensional space is called Pareto frontier \cite{Pareto2}. In general the computation of such an optimal set requires a time proportional to the cardinality $T_m$ \cite{godfrey}, that is a time exponential in the number $m$ of candidates.

Indeed also the computation of the Condorcet optimal consensus ranking with Kemeny distances (which is one element of the optimal set) is in general a NP-hard problem. However, if the $n$ ranked ballots given in input are not too dissimilar, such an optimum can be computed in polynomial time \cite{fixed-param}. Nonetheless the cases where our new criterion is meaningful are exactly those where the ranked ballots are not too similar.
We believe that for the computation of the optimal set in the large $m$ limit, one should resort to Monte Carlo methods, already successfully used in the computation of Kemeny optimal rankings \cite{MonteCarlo}.

\subsection*{Distribution of distances}

\begin{figure}[t]
\centering
\includegraphics[width=0.9\textwidth]{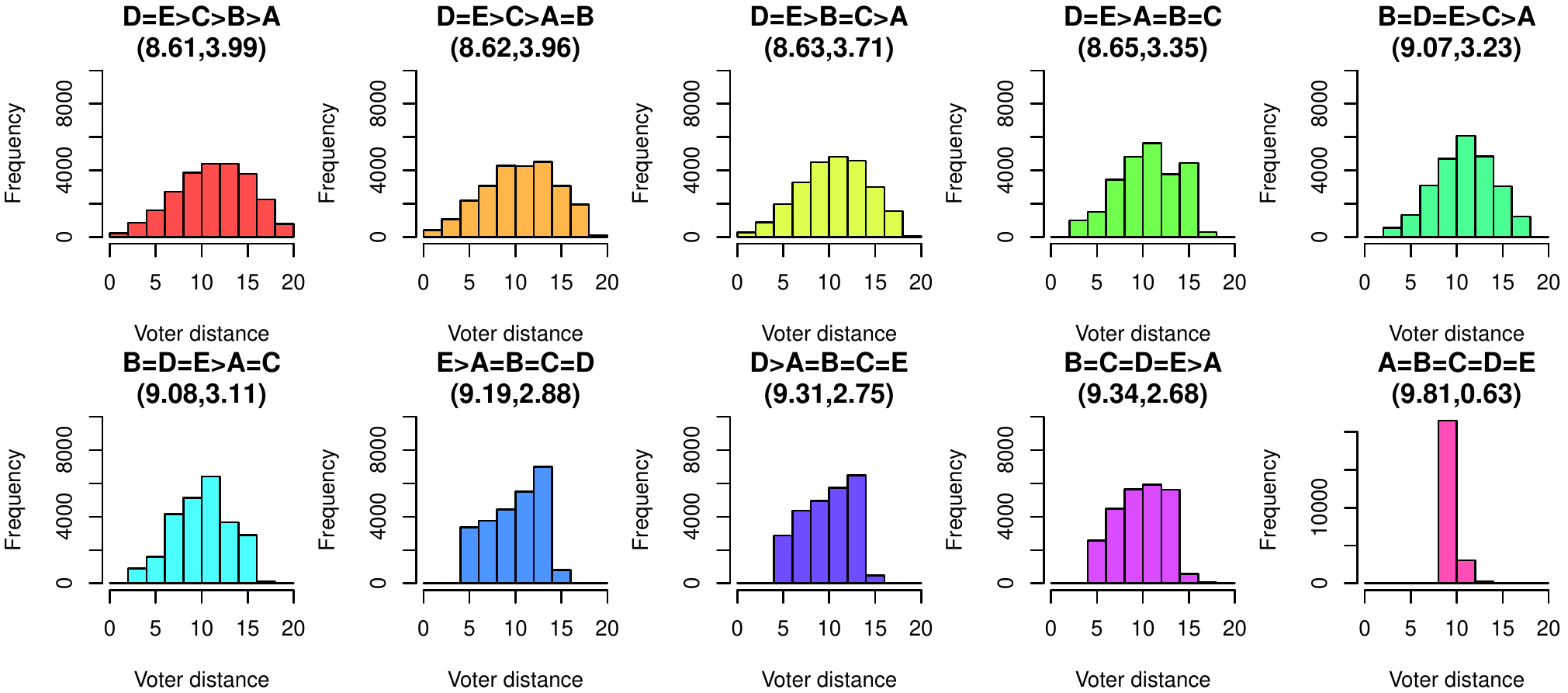}
\caption{For the jokes rating data discussed in the main paper, we show the distribution of distances to voter ballots for each of the rankings belonging to the optimal set. The title of each subplot indicates the ranking and its coordinates in the ($\mu$,$\sigma$) plane.}
\label{optimalHist}
\end{figure}

For the joke dataset discussed in the main paper, Figure \ref{optimalHist} presents a deeper analysis of the ten rankings belonging to the optimal set. These range from the Condorcet solution $\mathbf{c}^*$, with minimal $\mu$ but large $\sigma$, to the solution $\mathbf{\overline c}$ with minimum $\sigma$, but larger $\mu$. The figure shows the distribution of the distance between these rankings and the voter ballots (in other words distribution of voter dissatisfaction for these rankings). These distributions change from wide for the Condorcet ranking, where voter satisfaction is very uneven, to more narrow distributions as $\sigma$ decreases, and satisfaction becomes more comparable among voters.

It is worth noting that the consensus ranking $\mathsf{D\!=\!E\!>\!A\!=\!B\!=\!C}$ (fourth subplot on the first line in Figure \ref{optimalHist}) that we have proposed as the optimal one, based on the criterion of minimizing $\sigma(\mathbf{c})$ among rankings of low $\mu(\mathbf{c})$, is indeed the one that avoids strongly unsatisfied voters (distances larger than 15 are almost absent), without changing sensibly the mean value of the distribution.

\subsection*{Distance between rankings}

The Kemeny distance \cite{kemeny2} $d_\text{Kem}(\mathbf{r},\mathbf{s})$, that we have used in the main text, is one of the possible means of quantifying how dissimilar two rankings $\mathbf{r}$ and $\mathbf{s}$ are. Intuitively, the distance relates to how many pairwise comparisons of candidates do not match between the two rankings. For instance, if candidate $\mathsf{A}$ is preferred to candidate $\mathsf{B}$ in one ranking, but $\mathsf{B}$ is preferred to $\mathsf{A}$ in the other, that would count 1 in the distance. If one ranking considers $\mathsf{A\!=\!B}$ while the other does not, then that would count $1/2$ in the distance. By summing over all possible pairs, with $(\mathsf{A},\mathsf{B})$ and $(\mathsf{B},\mathsf{A})$ counted separately, one obtains the Kemeny distance between the two rankings. 

The computation of $d_\text{Kem}(\mathbf{r},\mathbf{s})$ is simpler if rankings are rewritten in terms of the score matrices $M(\mathbf{r})$:
\begin{equation}
M_{ij}(\mathbf{r}) = \left\{
\begin{array}{rl}
 1 & \text{if candidate $i$ is preferred to candidate $j$ in ranking $\mathbf{r}$}\\
-1 & \text{if candidate $j$ is preferred to candidate $i$ in ranking $\mathbf{r}$}\\
 0 & \text{otherwise}
\end{array}
\right.
\end{equation}
The Kemeny distance between rankings $\mathbf{r}$ and $\mathbf{s}$ is thus given by
\begin{equation}\label{kem}
d_\text{Kem}(\mathbf{r},\mathbf{s})=\frac{1}{2}\sum_{i,j}\left|M_{ij}(\mathbf{r})-M_{ij}(\mathbf{s})\right|
\end{equation}

\subsection*{Some winner selection methods}

In the plots in the main paper we have shown the consensus rankings obtained by some well-known winner selection methods, Copeland, Schulze and Tideman \cite{bookVoting2}. Here we provide a detailed description of these methods, which are the most commonly used in situations where the voter ballots are lists of candidates ordered by preference (ranked ballots).

We consider the same situation as in the main paper, where $n$ voters express their preferences about $m$ candidates. The ballot of each voter can be conveniently mapped in a vector $\mathbf{r}$ of $m$ integers representing the positions of each candidate in the preference list. For example the ballot $\mathsf{C\!>\!A\!>\!E\!>\!D\!>\!B}$ corresponds to the vector $\mathbf{r} = (2,5,1,4,3)$. From the $n$ vectors $\mathbf{r}^{(k)}$, with $k=1\ldots,n$,  representing the voter ballots we can build the matrix of total preferences whose elements are
\[
P_{ij} = \sum_{k=1}^n \mathbb{I}(r^{(k)}_j > r^{(k)}_i)\;,
\]
where the indicator function $\mathbb{I}$ is defined as
\[
\mathbb{I}({\rm condition}) = \left\{
\begin{array}{rl}
1 & {\rm if\ condition\ is\ true}\\
0 & {\rm if\ condition\ is\ false}
\end{array}
\right.
\]
In practice the matrix element $P_{ij}$ counts how many voters prefer candidate $i$ to candidate $j$. The result of any voting method based only on pairwise comparisons between candidates can be obtained from matrix $P$.


A method of selecting a consensus ranking based on scores is \emph{Copeland}, also known as the \emph{pairwise comparison}. Candidates are ranked according to the score $C_i$ that counts the number of pairwise comparisons won plus half of those tied
\[
C_i = \sum_{j=1}^m \left[\mathbb{I}(P_{ij} > P_{ji}) + \frac12 \mathbb{I}(P_{ij} = P_{ji})\right]\;.
\]
The Copeland candidate(s) is the one with maximum $C_i$.

The \emph{Schulze} method is also based on pairwise comparisons between candidates. To compute the Schulze ranking from the matrix $P$ we first have to compute the matrix $B$ of \emph{beatpaths}, by initialing it as $B_{ij} = P_{ij}$ and then iterating until convergence
\[
B_{ij} = \max\left( B_{ij},\;\max_k \min(B_{ik}, B_{kj}) \right)\;.
\]
The number of iterations to make the matrix $B$ converge is given by the length of the longest beatpath, which is at most the number of candidates $m$.
Successively, candidates are ranked according to a score similar to the pairwise one for the $B$ matrix, that is
\[
Z_i = \sum_{j=1}^m \left[ \mathbb{I}(B_{ij} > B_{ji}) + \frac12 \mathbb{I}(B_{ij} = B_{ji})\right]\;.
\]
The Schulze candidate(s) is the one with maximum $Z_i$.

\emph{Tideman} is a further method of selecting a consensus ranking. To compute the Tideman solution, the elements of matrix $P$ are sorted in a decreasing order and taken into account one by one. When element $P_{ij}$ is considered, the relative order $i > j$ in the final ranking is assigned unless in contrast with the partial ordering already fixed by larger values of $P$ previously considered.

\subsection*{Statistics of the raw data}

\begin{figure}
\centering
\includegraphics[width=0.8\textwidth]{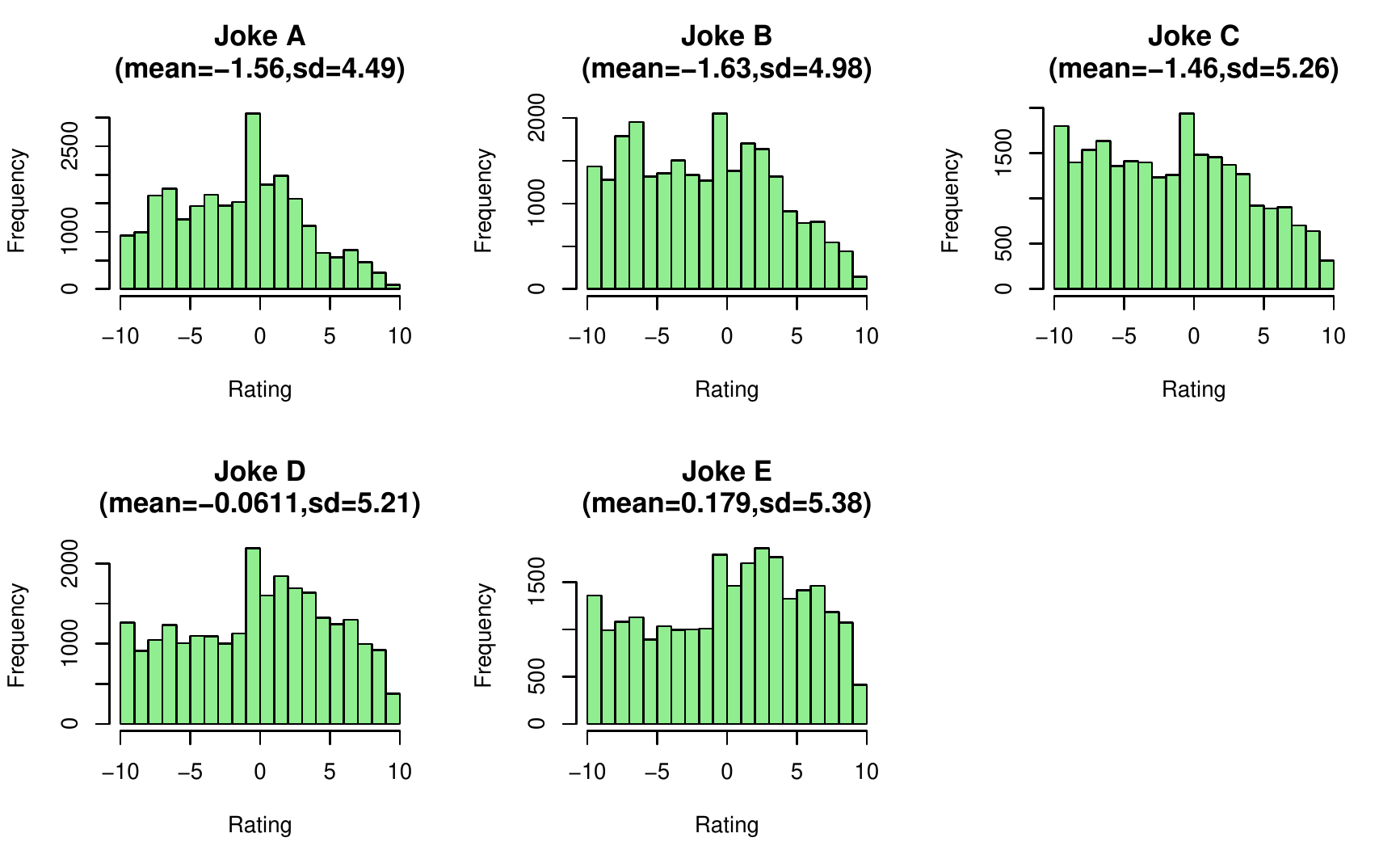}
\caption{Distribution of ratings for the five jokes discussed in the main paper. Mean and standard deviation (sd) values of the ratings are also shown for each joke.}
\label{jokeHist}
\end{figure}

\begin{figure}
\centering
\includegraphics[width=0.8\textwidth]{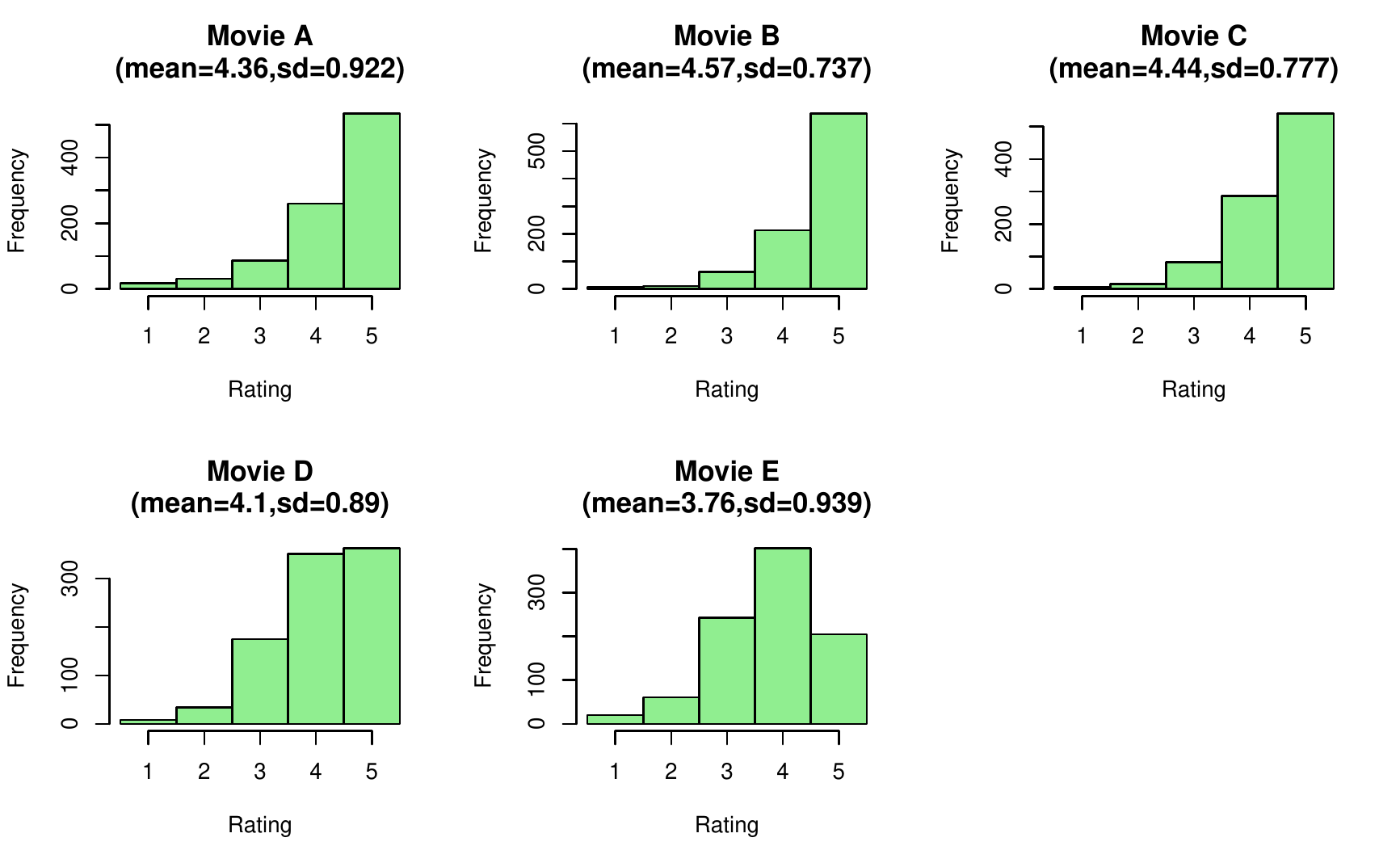}
\caption{Distribution of ratings for the five movies discussed in the main paper. Mean and standard deviation (sd) values of the ratings are also shown for each movie.}
\label{movieHist}
\end{figure}

In order to provide a better view over the datasets used in the main paper, we include here the distributions of rating values for the five jokes and five movies analyzed (Figures \ref{jokeHist} and \ref{movieHist}, respectively).

\begin{table}
\centering
\begin{tabular}{ |c |c |}
\hline
Number of voters&Ballot value\\
\hline
1 & $\mathsf{E\!>\!C\!>\!D\!>\!A\!=\!B}$ \\ 
\hline
1 & $\mathsf{E\!>\!A\!=\!B\!=\!C\!=\!D}$ \\ 
\hline
1 & $\mathsf{C\!=\!D\!=\!E\!>\!A\!=\!B}$ \\ 
\hline
1 & $\mathsf{C\!=\!E\!>\!D\!>\!A\!=\!B}$ \\ 
\hline
1 & $\mathsf{D\!>\!E\!>\!C\!>\!A\!=\!B}$ \\ 
\hline
1 & $\mathsf{D\!>\!C\!=\!E\!>\!A\!=\!B}$ \\
\hline
1 & $\mathsf{E\!>\!C\!>\!B\!>\!D\!>\!A}$ \\ 
\hline
1 & $\mathsf{C\!=\!D\!>\!A\!=\!B\!=\!E}$ \\
\hline
1 & $\mathsf{C\!>\!B\!=\!E\!>\!A\!>\!D}$ \\ 
\hline
1 & $\mathsf{C\!>\!A\!>\!B\!=\!E\!>\!D}$ \\
\hline
1 & $\mathsf{C\!>\!D\!>\!E\!>\!A\!=\!B}$ \\ 
 \hline
1 & $\mathsf{C\!>\!B\!>\!E\!>\!D\!>\!A}$ \\
\hline
2 & $\mathsf{A\!=\!B\!=\!C\!=\!D\!=\!E}$ \\ 
\hline
\end{tabular}
\caption{Ballot values for the AIRESIS data discussed in the main paper.}\label{airesisBallot}
\end{table}

Joke data have wide distributions for the ratings of each joke, with jokes D and E having right-skewed distributions unlike the other jokes, thus explaining their supremacy in the consensus rankings provided by different winner selection methods (see discussion of Figure \ref{joke} in the main text). 
It is worth noting that, although mean ratings for jokes D and E are larger than mean ratings for jokes A, B and C, any other difference among the means is not significative, given that rating standard deviations are of order 5.

All the five movies we have considered have very high ratings (see Figure~\ref{movieHist}), thus explaining the results obtained in Figure~\ref{movie} in the main text.

For the AIRESIS dataset analyzed in the main paper, we include directly the votes expressed by users, since a small number of ballots are present here. Table \ref{airesisBallot} shows the number of voters opting for a particular ranked ballot.

\begin{figure}[t]
\centering
\fbox{\includegraphics[width=0.7\textwidth]{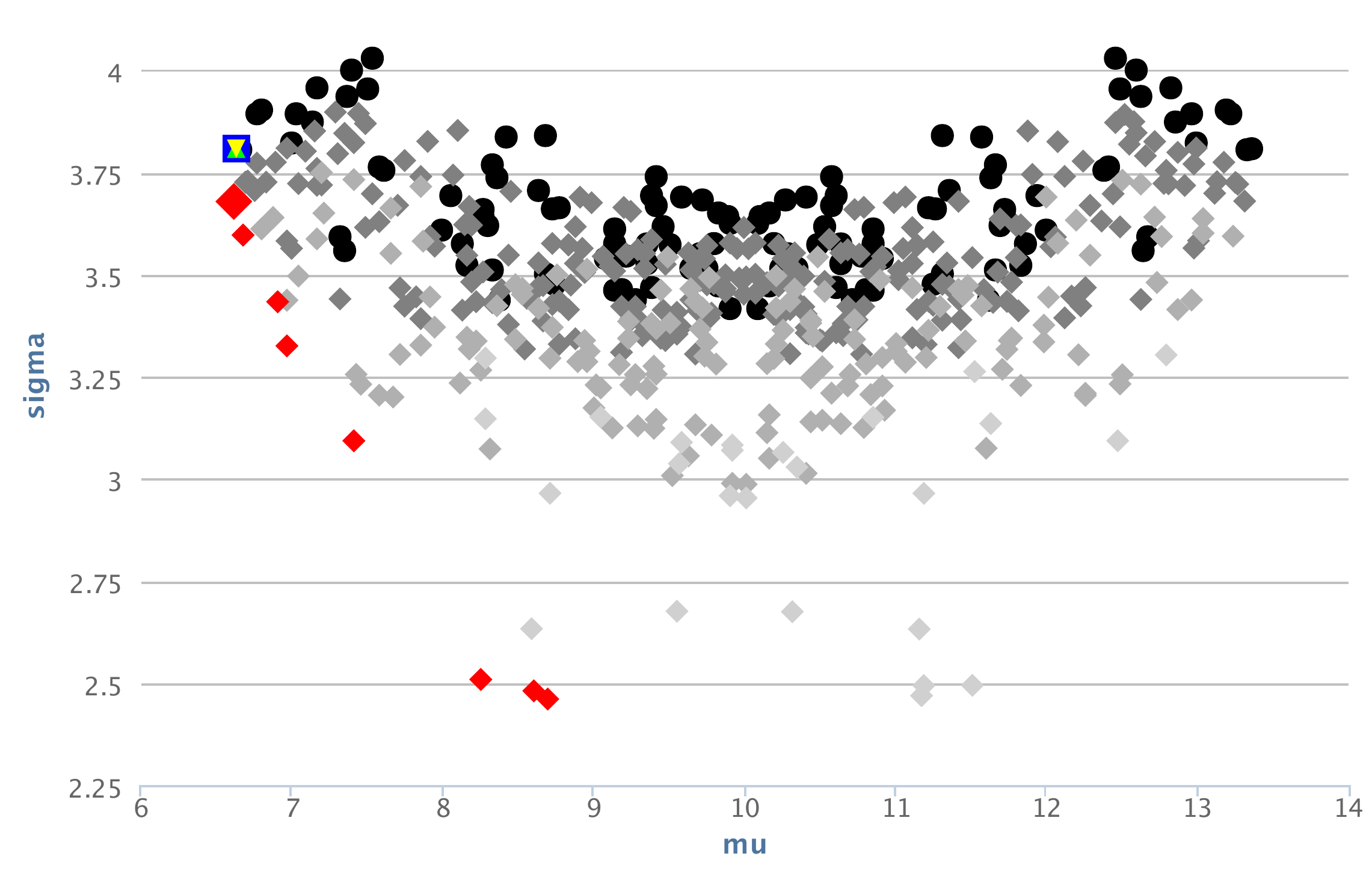}}
\\
\hspace{1cm}{\large$\bullet$}  Rankings with no ties \hfill
{\color{red}\large$\vardiamond$} Condorcet solution ($\mathsf{C\!>\!D\!>\!A\!>\!B\!=\!E}$)\\
\hspace{1cm}{\color{grey1}$\vardiamond$} Rankings with one tie  \hfill
{\color{blue}$\blacksquare$} Copeland solution ($\mathsf{C\!>\!D\!>\!A\!>\!E\!>\!B}$)\\
\hspace{1cm}{\color{grey2}$\vardiamond$} Rankings with two ties \hfill
{\color{yellow}$\blacktriangledown$} Schulze solution ($\mathsf{C\!>\!D\!>\!A\!>\!E\!>\!B}$)\\
\hspace{1cm}{\color{grey3}$\vardiamond$} Rankings with three ties \hfill
{\color{green}\hspace{1cm}$\blacktriangle$} Tideman solution ($\mathsf{C\!>\!D\!>\!A\!>\!E\!>\!B}$)\\
\hspace{1cm}{\color{grey4}$\vardiamond$} Rankings with four ties \hfill 
{\color{red}\footnotesize$\vardiamond$ }Optimal set
\caption{One more example from the jokes rating data, where a different set of 5 jokes rated by 16049 users is analyzed. Again, for visualization purposes, the solution $\mathsf{A\!=\!B\!=\!C\!=\!D\!=\!E}$ at (9.84,0.46) has been omitted.}
\label{joke1}
\end{figure}

\subsection*{Further results on real polls}

Due to space restrictions, the discussion in the main paper was reduced to only a few examples of vote instances. However the new method of analysis we have introduced seems to work nicely on any voting instance we have studied. Here we report some more results from jokes and Airesis data in the usual two-dimensional space $(\mu,\sigma)$.

Figure \ref{joke1} plots $\sigma$ and $\mu$ for another set of five jokes from the Jester database, rated by 16049 voters. As before, for each voter the ranked ballot is obtained by ordering the five jokes based on the rating the voter provided for them. The figure shows that standard methods obtain good results in terms of $\mu$, however $\sigma$ can be improved by selecting a different consensus ranking, which in this case contains equalities.

\begin{figure}[t]
\centering
\fbox{\includegraphics[width=0.48\textwidth]{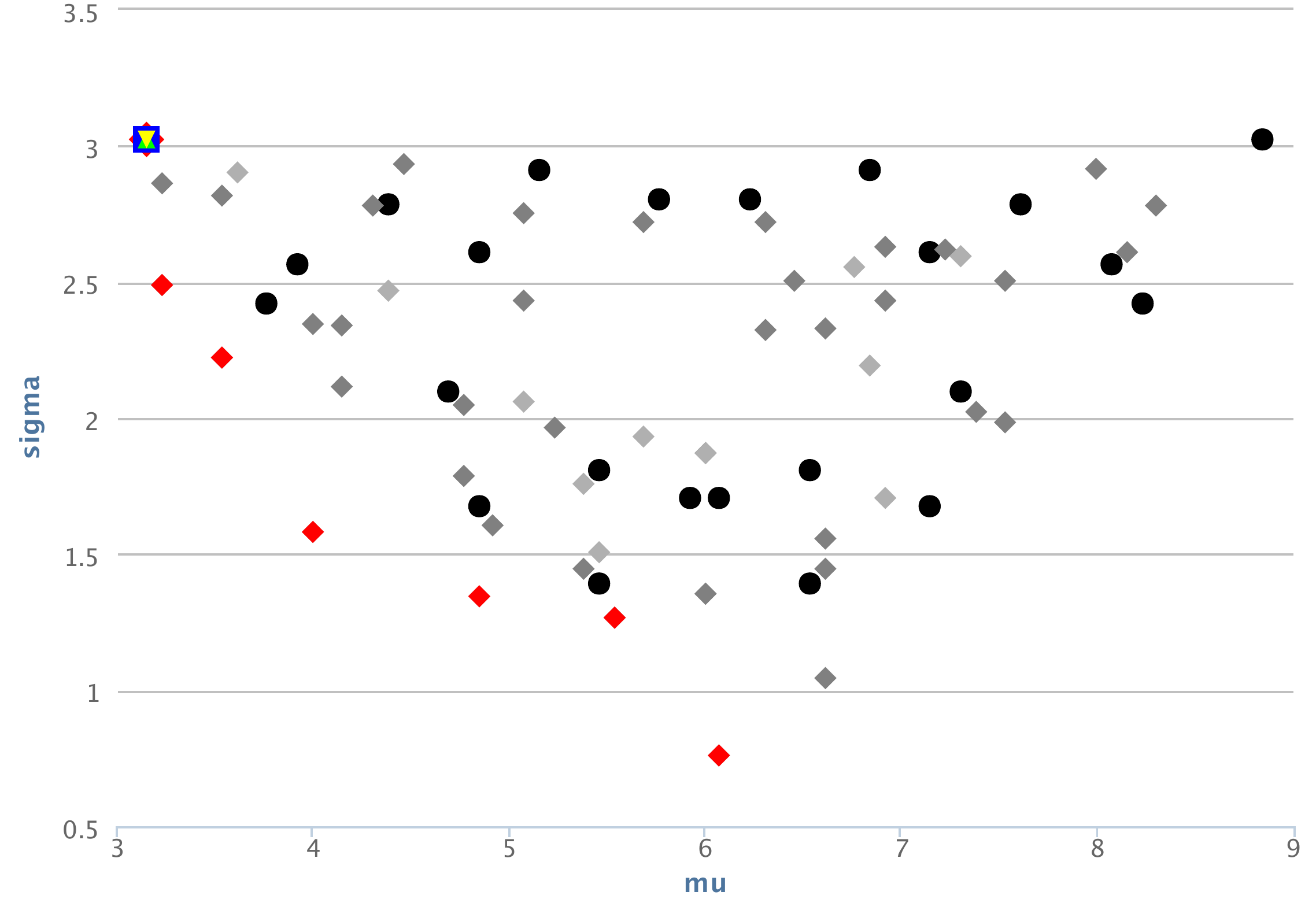}}
\fbox{\includegraphics[width=0.48\textwidth]{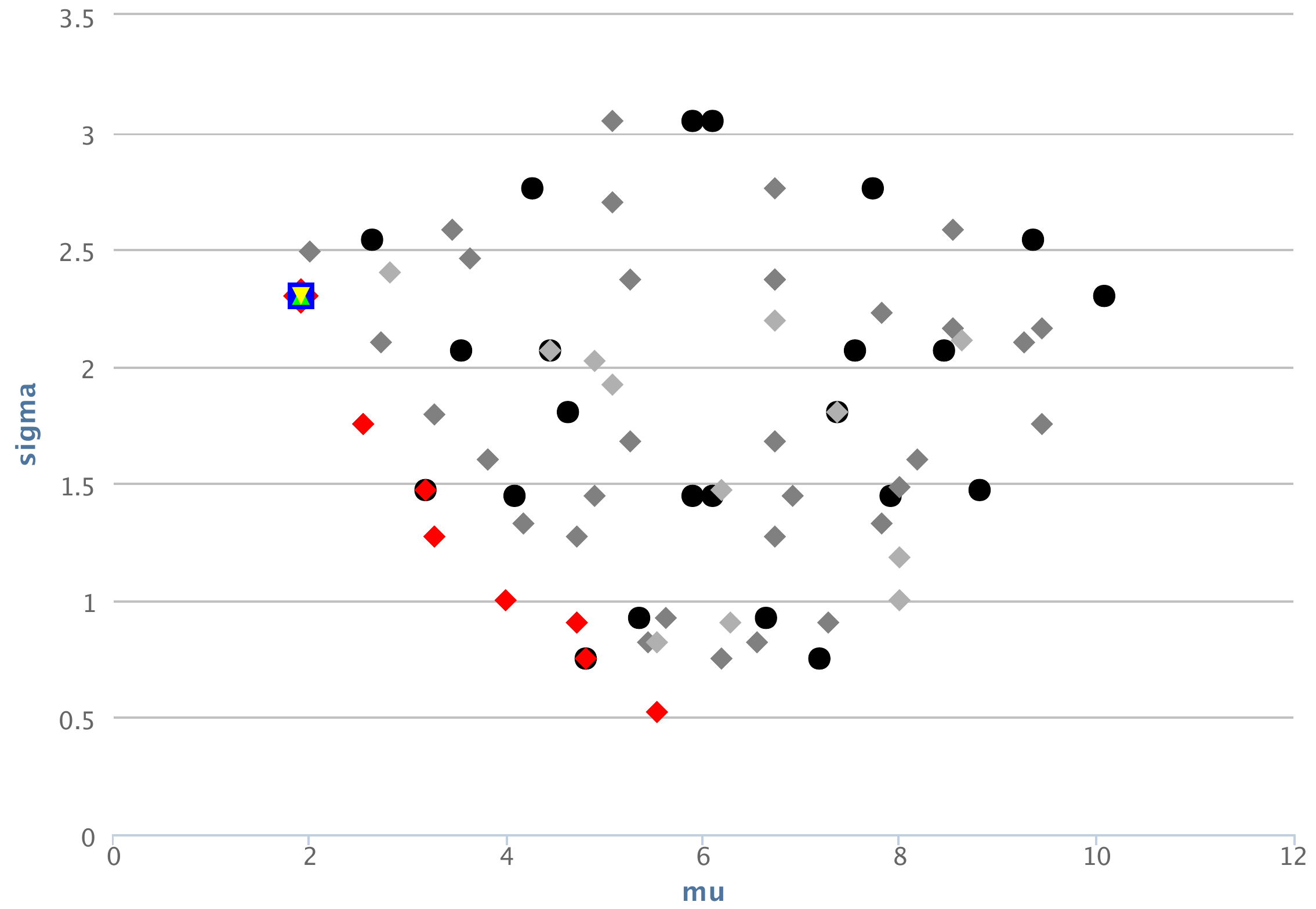}}
\fbox{\includegraphics[width=0.48\textwidth]{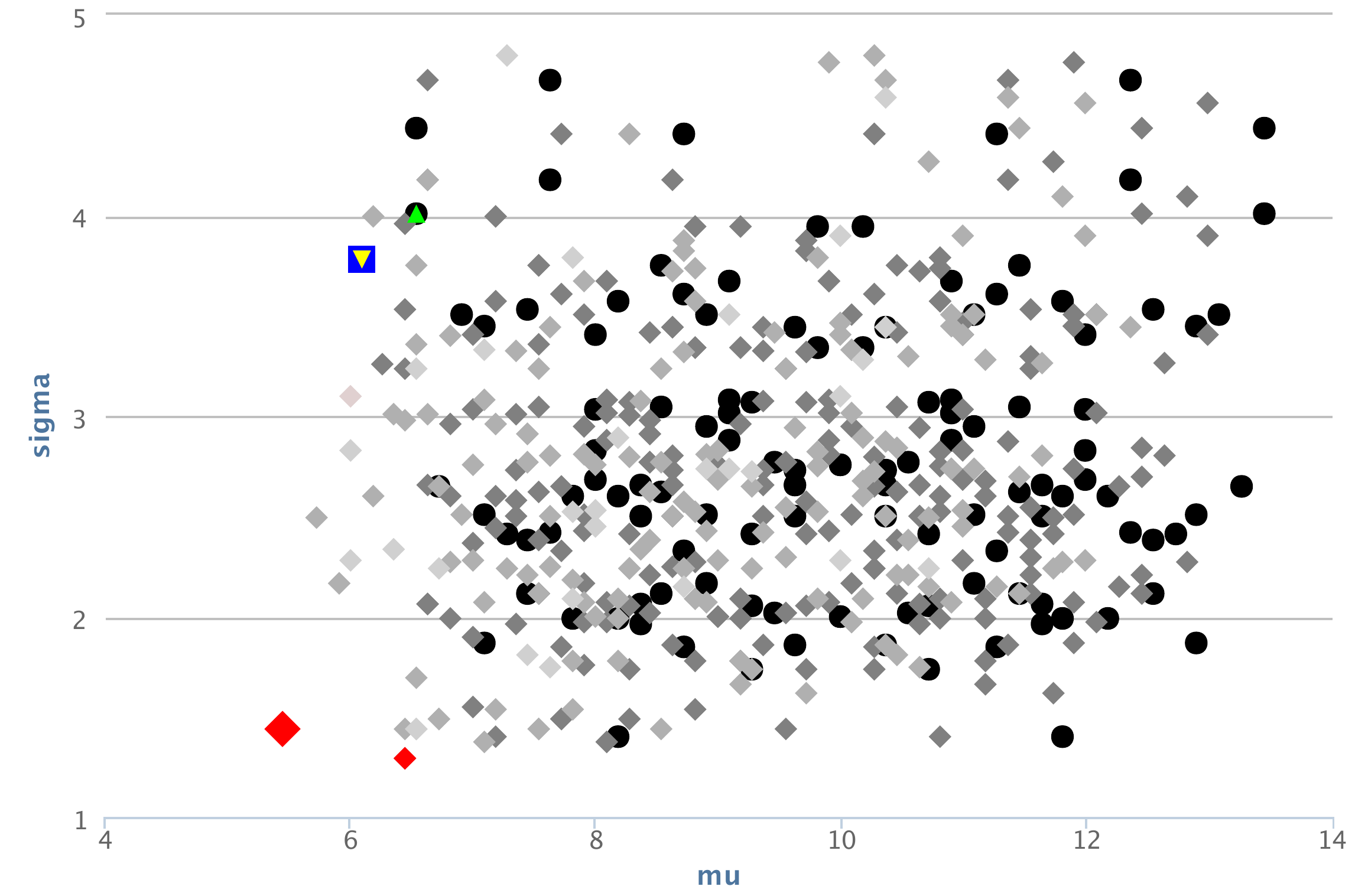}}
\\
\hspace{1cm}{\large$\bullet$}  Rankings with no ties \hfill
{\color{red}\large$\vardiamond$} Condorcet solution \\
\hspace{1cm}{\color{grey1}$\vardiamond$} Rankings with one tie  \hfill
{\color{blue}$\blacksquare$} Copeland solution \\
\hspace{1cm}{\color{grey2}$\vardiamond$} Rankings with two ties \hfill
{\color{yellow}$\blacktriangledown$} Schulze solution  \\
\hspace{1cm}{\color{grey3}$\vardiamond$} Rankings with three ties \hfill
{\color{green}\hspace{1cm}$\blacktriangle$} Tideman solution \\
\hspace{1cm}{\color{grey4}$\vardiamond$} Rankings with four ties \hfill 
{\color{red}\footnotesize$\vardiamond$ }Optimal set
\caption{Further examples from polls run on the AIRESIS platform.}\label{airesis1}
\end{figure}

Further examples of data from the AIRESIS platform are shown in Figure \ref{airesis1}. These show again that, in general, existing winner selection methods are not optimal with respect to minimizing $\sigma(\mathbf{c})$ (and sometimes neither in minimizing $\mu(\mathbf{c})$). A clear improvement can be achieved by considering rankings in the optimal set we have introduced.

\begin{figure}[t]
\centering
\includegraphics[width=0.8\textwidth]{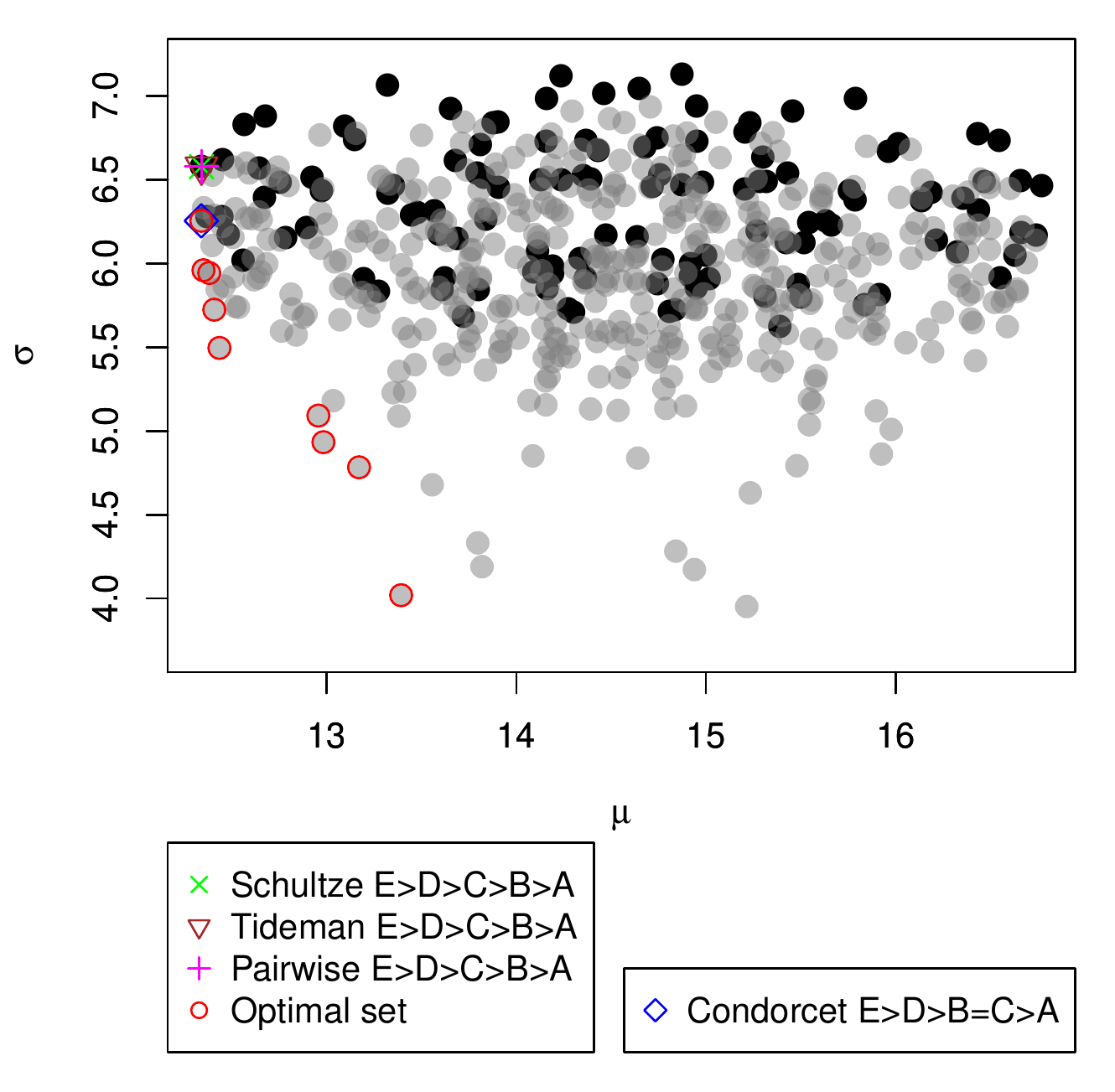}
\caption{Example of using an alternative distance for the same jokes rating data shown in Figure~\ref{joke} of the main paper. The distance used here is Kemeny with position weights. The structure of the ranking space remains the same.}
\label{weighted}
\end{figure}

\subsection*{Different distance measures}

The new method proposed in this paper has been illustrated using the Kemeny distance among rankings. 
There are of course many different distances to be considered and the choice among them depends mostly on the application.
There are classes of distances bounded from above and below by multiplicative constants \cite{digr}.

An important generalization of a given distance is to introduce weights. In some  problems there are, for instance, candidates of different importance and it is possible to introduce \emph{candidate weights}. In other problems the position where the candidate is in the ballot counts, e.g.\ candidates near the top of the ballot are more likely to be important than those at the bottom, so it may be useful to introduce \emph{position weights}. Weights can be introduced also by dividing candidates in homogeneity classes: i.e.\ swaps of two candidates belonging to the same class have a smaller weight than swaps between different classes. The properties of these weighted distances are studied in \cite{kmvs}.

Moreover, beyond the distances in $\ell_1$ metrics, like the Kemeny one, it is natural to consider also distances in $\ell_2$ metrics \cite{cs} or even higher order metrics.

We have checked several of the above generalizations and found that the general structure of the
rankings in the $(\mu,\sigma)$ space, and hence the conclusions drawn in the main paper, does not change with the distance measure. As a representative example we show in Figure~\ref{weighted} the same data from the joke ratings presented in Figure~\ref{joke} of the main paper, but using a Kemeny distance modified with position weights. This gives more importance to candidates at the top of a voter ballot (a pair swapped at the top counts more in the distance than a pair swapped at the bottom of the ranking). Specifically, in the computation of the distance, equation  \ref{kem} is replaced by:
\begin{equation}
d_\text{Kem,pos}(\mathbf{r},\mathbf{s})=\frac{1}{2}\frac{m}{4m-6}\sum_{i,j}{(\left|M_{ij}(\mathbf{r})-M_{ij}(\mathbf{s})\right|(2m-r_i-r_j))}
\end{equation}
The structure of the ranking space is mostly unchanged and indeed the consensus ranking that we identify through the minimization of $\sigma(\mathbf{c})$ among the low $\mu(\mathbf{c})$ rankings is still the same one we found with the standard Kemeny distance, i.e.\ $\mathsf{D\!=\!E\!>\!A\!=\!B\!=\!C}$. 

Another possible generalization is to measure the fluctuations in the voters' satisfaction, that is the $\sigma(\mathbf{c})$ in formula (\ref{std}), not only through the measure of the standard deviation of the distances, but also by the average of the absolute deviations from the mean. Since the two measures are both robust estimators of the fluctuations, we expect them to give similar results when the number of voters is large.

\subsection*{Noise sensitivity and how to estimate the uncertainty on $\mu$ and $\sigma$}

A very important feature of a fair voting method is the robustness with respect to small noise fluctuations, that is one would like to avoid a consensus ranking changing drastically because of some noise induced fluctuations.
When running a poll, there may be several different sources of noise, the simplest one being fluctuations in the number of participants.
In Fig.~\ref{stab} in the main paper we have shown fluctuations in $\mu$ and $\sigma$ induced by a random elimination of a fraction $\alpha$ of votes, uniformly chosen among the $n$ votes available. Data shown in Fig.~\ref{stab} have been obtained with $\alpha=0.1$ and $\alpha=0.2$, and they show a rather good collapse once scaled according to the following scaling.

Given a set of $n$ votes, where ranking $\mathbf{r}$ appears $n(\mathbf{r})$ times, the subsampling produces a subset of $(1-\alpha)n$ votes with frequencies $k(\mathbf{r})$ Binomially  distributed in $[0,n(\mathbf{r})]$, with mean ${\mathbb E} [k(\mathbf{r})]=(1-\alpha) n(\mathbf{r})$ and variance ${\mathbb V} [k(\mathbf{r})]=\alpha (1-\alpha) n(\mathbf{r})$.
Calling $\mu(\mathbf{c})$ the mean Kemeny distance computed with frequencies $n(\mathbf{r})$
and $\mu_\alpha(\mathbf{c})$ those computed with frequencies $k(\mathbf{r})$, a simple computation based on the assumption of independence of the $k(\mathbf{r})$, shows that the variance ${\mathbb V}[\mu_\alpha]$ over many subsamplings scales proportionally to $\alpha/(1-\alpha)$.
In other words the quantity ${\mathbb V}[\mu_\alpha](1-\alpha)/\alpha$ should be roughly $\alpha$-independent (as shown in Fig.~\ref{stab}) and is closely related to the uncertainty on $\mu(\mathbf{c})$.

\begin{figure}
\centering
\includegraphics[width=0.73\textwidth]{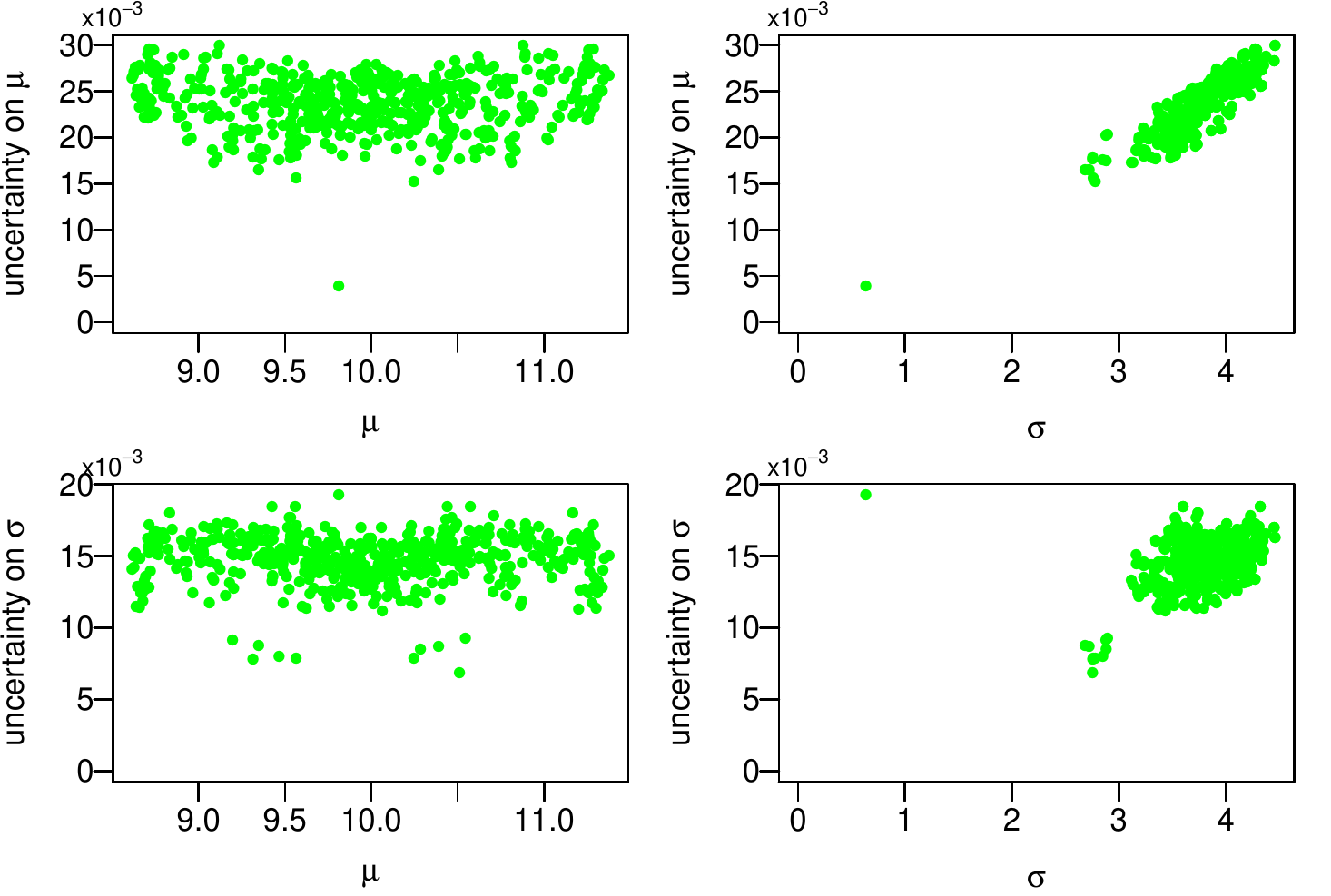}
\caption{Uncertainties on $\mu(\mathbf{c})$ and $\sigma(\mathbf{c})$, as obtained from the bootstrap analysis, are very close to noise fluctuations shown in Fig.~\ref{stab}, and make consensus rankings of low $\sigma$ more reliable.}
\label{bootstrap}
\end{figure}


A standard way to compute the uncertainty on any experimental measure is the so-called {\it bootstrap} method of analysis: starting from the original data set of $n$ measures, new data sets of $n$ measures each can be obtained by randomly choosing the original measures with repetitions (that is each measure appears in a new data set a number of times which is a Poisson random variable of mean 1). For any observable, e.g. $\mu(\mathbf{c})$, computed on the original data set, the uncertainty can be computed from its fluctuations among the new data sets.
In Fig.~\ref{bootstrap} we show uncertainties for $\mu(\mathbf{c})$ and $\sigma(\mathbf{c})$, and we notice they are very close to fluctuations induced by subsampling (shown in Fig.~\ref{stab}).


\end{document}